\documentclass[12pt]{iopart}
\usepackage{latexsym}
\usepackage{amscd,amsfonts}
\usepackage[dvips]{graphics,graphics,color}
\begin{document}
\definecolor{r}{rgb}{1,0,0}
\definecolor{b}{rgb}{0,0,1}
\definecolor{g}{cmyk}{0,0,1,0}
\jl{1}

\jl{1}
 \def\lambdabar{\protect\@lambdabar}
\def\@lambdabar{%
\relax
\bgroup
\def\@tempa{\hbox{\raise.73\ht0
\hbox to0pt{\kern.25\wd0\vrule width.5\wd0
height.1pt depth.1pt\hss}\box0}}%
\mathchoice{\setbox0\hbox{$\displaystyle\lambda$}\@tempa}%
{\setbox0\hbox{$\textstyle\lambda$}\@tempa}%
{\setbox0\hbox{$\scriptstyle\lambda$}\@tempa}%
{\setbox0\hbox{$\scriptscriptstyle\lambda$}\@tempa}%
\egroup
}

\def\bbox#1{%
\relax\ifmmode
\mathchoice
{{\hbox{\boldmath$\displaystyle#1$}}}%
{{\hbox{\boldmath$\textstyle#1$}}}%
{{\hbox{\boldmath$\scriptstyle#1$}}}%
{{\hbox{\boldmath$\scriptscriptstyle#1$}}}%
\else
\mbox{#1}%
\fi
}
\newcommand{\muv}{\bbox{\mu}}
\newcommand{\mc}{{\mathcal M}}
\newcommand{\pc}{{\mathcal P}}
\newcommand{\mct}{\bbox{\mathcal M}}
\newcommand{\pct}{\bbox {\mathcal P}}
\newcommand{\fsf}{{\sf F}}
\newcommand{\fsft}{\bbox{{\sf F}}}
 \newcommand{\mv}{\bbox{m}}
\newcommand{\pv}{\bbox{p}}
\newcommand{\tv}{\bbox{t}}
\def\msf{\hbox{{\sf M}}}
\def\msft{\bbox{{\sf M}}}
\def\psf{\hbox{{\sf P}}}
\def\psft{\bbox{{\sf P}}}
\def\Nsf{\hbox{{\sf N}}}
\def\Nsft{\bbox{{\sf N}}}
\def\Tsf{{\sf T}}
\def\Tsft{\bbox{{\sf T}}}
\def\Asf{\hbox{{\sf A}}}
\def\Asft{\bbox{{\sf A}}}
\def\Bsf{\hbox{{\sf B}}}
\def\Bsft{\bbox{{\sf B}}}
\def\Lsf{\hbox{{\sf L}}}
\def\Lsft{\bbox{{\sf L}}}
\def\Ssf{\hbox{{\sf S}}}
\def\Ssft{\bbox{{\sf S}}}
\def\Mtens{\bi{M}}
\def\msfsim{\bbox{{\sf M}}_{\scriptstyle\rm(sym)}}
\newcommand{\mcsim}{ {\sf M}_{ {\scriptstyle \rm {(sym)} } i_1\dots i_n}}
\newcommand{\mcs}{ {\sf M}_{ {\scriptstyle \rm {(sym)} } i_1i_2i_3}}

\newcommand{\beqan}{\begin{eqnarray*}}
\newcommand{\eeqan}{\end{eqnarray*}}
\newcommand{\beqa}{\begin{eqnarray}}
\newcommand{\eeqa}{\end{eqnarray}}

 \newcommand{\suml}{\sum\limits}
 \newcommand{\sumd}{\suml_{\mathcal D}}
\newcommand{\intl}{\int\limits}
\newcommand{\rvec}{\bbox{r}}
\newcommand{\xivec}{\bbox{\xi}}
\newcommand{\Avec}{\bbox{A}}
\newcommand{\Rvec}{\bbox{R}}
\newcommand{\Evec}{\bbox{E}}
\newcommand{\Bvec}{\bbox{B}}
\newcommand{\Svec}{\bbox{S}}
\newcommand{\avec}{\bbox{a}}
\newcommand{\nablav}{\bbox{\nabla}}
\newcommand{\nuvec}{\bbox{\nu}}
\newcommand{\bvec}{\bbox{\beta}}
\newcommand{\vvec}{\bbox{v}}
\newcommand{\jvec}{\bbox{J}}
\newcommand{\nvec}{\bbox{n}}
\newcommand{\pvec}{\bbox{p}}
\newcommand{\mvec}{\bbox{m}}
\newcommand{\evec}{\bbox{e}}
\newcommand{\eps}{\varepsilon}
\newcommand{\la}{\lambda}
\newcommand{\rad}{\mbox{\footnotesize rad}}
\newcommand{\scr}{\scriptstyle}
\newcommand{\latens}{\bbox{\sf{\Lambda}}}
\newcommand{\lasf}{{\sf{\Lambda}}}
\newcommand{\pitens}{\sf{\Pi}}
\newcommand{\cm}{{\cal M}}
\newcommand{\cp}{{\cal P}}
\newcommand{\beq}{\begin{equation}} 
\newcommand{\eeq}{\end{equation}}
\newcommand{\ptens}{\bbox{\sf{P}}}
\newcommand{\Ptens}{\bbox{P}}
\newcommand{\Ttens}{\bbox{\sf{T}}}
\newcommand{\Ntens}{\bbox{\sf{N}}}
\newcommand{\Ncal}{\bbox{{\cal N}}}
\newcommand{\Atens}{\bbox{\sf{A}}}
\newcommand{\Btens}{\bbox{\sf{B}}}
\newcommand{\dom}{\mathcal{D}}
\newcommand{\al}{\alpha}
\newcommand{\sym}{\scriptstyle \rm{(sym)}}
\newcommand{\Tcal}{\bbox{{\mathcal T}}}
\newcommand{\Nmc}{{\mathcal N}}
\renewcommand{\d}{\partial}
\def\rmi{{\rm i}}
\def\rme{\hbox{\rm e}}
\def\rmd{\hbox{\rm d}}
\newcommand{\ct}{\mbox{\Huge{.}}}
\newcommand{\Laop}{\bbox{\Lambda}}
\newcommand{\Ssfs}{{\scriptstyle \Ssft^{(n)}}}
\newcommand{\Lsfs}{{\scriptstyle \Lsft^{(n)}}}
\newcommand{\psfr}{\widetilde{\psf}}
\newcommand{\msfr}{\widetilde{\msf}}
\newcommand{\msftr}{\widetilde{\msft}}
\newcommand{\psftr}{\widetilde{\psft}}
\newcommand{\pvr}{\widetilde{\pvec}}
\newcommand{\mvr}{\widetilde{\mvec}}
\newcommand{\qdot}{\stackrel{\cdot\cdot\cdot\cdot}}
\newcommand{\bsy}{\hbox}
\newcommand{\ointl}{\oint\limits}
\newcommand{\pisf}{{\sf \Pi}}
\newcommand{\ssft}{\bbox{{\sf S}}}
\newcommand{\ssf}{{\sf S}}
\def\Nsf{{\sf N}}
\newcommand{\gamsf}{{\sf \Gamma}}
\newcommand{\gamsft}{\bsy{\sf \Gamma}}
\newcommand{\mlrt}{\stackrel{\leftrightarrow}{\msft}}
\newcommand{\mlr}{\stackrel{\leftrightarrow}{\msf}}
\newcommand{\ab}{\v{a}} 
\newcommand{\ai}{\^{a}} 
\newcommand{\ib}{\^{\i}} 
\newcommand{\tb}{\c{t}} 
\newcommand{\st}{\c{s}}
\newcommand{\Ab}{\v{A}} 
\newcommand{\Ai}{\^{A}} 
\newcommand{\Ib}{\^{I}} 
\newcommand{\Tb}{\c{T}}
\newcommand{\St}{\c{S}}
\newcommand{\lavec}{\bbox{\Lambda}}
\newcommand{\esft}{\bbox{{\sf e}}}

\title{On the $\delta$-singularities of the    electromagnetic field }
\author{C.\ Vrejoiu\footnote{E-mail :  vrejoiu@fizica.unibuc.ro}, R.\ Zus \footnote{E-mail: roxana.zus@fizica.unibuc.ro}}
\address{University of Bucharest, Department of Physics,  \\
PO Box MG - 11, Bucharest-Magurele, RO - 077125,
 Romania }

\begin{abstract}
The singularities of the electromagnetic field are derived to include all the point-like multipoles representing an electric charge and current distribution. We show that for higher orders, it is more efficient to have fields represented in terms of symmetric and trace free moments. 
In the static case, the delta-singularities are expressed for arbitrary multipole orders, while in the dynamic case we restrict ourselves to the lower orders. The algorithm we give can be easily extended to the next orders.
\end{abstract}
\section{Introduction}\label{intro}\footnote{This paper is a revisited, completed and finalized version of two previous  articles: arXiv:0912.4684 and arXiv:1001.4114}
In the cases of electrostatic and magnetostatic fields of point-like dipoles, one has the well-known procedure of introducing  Dirac $\delta$-function terms for obtaining correct expressions of the electric and magnetic fields defined on the entire space.  The corresponding field expressions take the following form \cite{Jackson}: 
\beqa\label{1.1}
\fl\;\;\;\;\;\;\Evec_{\pvec}(\rvec)=-\frac{1}{3\eps_0}\,\pvec\,\delta(\rvec)+\frac{1}{4\pi\eps_0}\left(\frac{3(\nuvec\cdot\pvec)\nuvec-\pvec}{r^3}\right)_{r\ne 0}=-\frac{1}{3\eps_0}\,\pvec\,\delta(\rvec) + \big(\Evec\big)_{r\ne 0}\ ,
\eeqa
where $\nuvec=\rvec/r$, and
\beqa\label{1.2}
\fl\;\;\;\;\;\;\Bvec_{\mvec}(\rvec)= \frac{2\mu_0}{3}\,\mvec\,\delta(\rvec)+\frac{\mu_0}{4\pi}\left(\frac{3(\nuvec\cdot\mvec)\nuvec-\mvec}{r^3}\right)_{r\ne 0}=
\frac{2\mu_0}{3}\,\mvec\,\delta(\rvec)+ \big(\Bvec\big)_{r\ne 0}\ .
\eeqa
In these equations, by $(\dots)_{r\ne 0}$ we understand an expression in which the derivatives are calculated supposing $r\ne 0$, representing some well-known expressions of the fields.
The expressions from equations \eref{1.1} and \eref{1.2} are introduced in Ref.\  \cite{Jackson} as  conditions of compatibility with the average value of the  electric or magnetic field  over a spherical domain containing all the charges or currents inside. Another procedure for introducing equations \eref{1.1} and \eref{1.2} is based on an extension of the derivative $\d_i\d_j/(1/r)$ to the entire space \cite{Frahm}:
\beqa\label{1.3}
\d_i\d_j\frac{1}{r}=\left(\frac{3\,\nu_i\nu_j\,-\,\delta_{ij}}{r^3}\right)_{r\ne 0}\,-\frac{4\pi}{3}\,\delta_{ij}\,\delta(\rvec)
\eeqa
and this procedure will be applied in the present paper.
A more pedagogical and suitable approach for understanding the origin of the difference between  the electric and magnetic cases  is done in Ref.\  \cite{Leung06}.  Refs.\  \cite{Werner} and \cite{Leung07}  contain  generalizations of the equations \eref{1.1} and \eref{1.2} to the dynamic case for  oscillating electric and magnetic dipoles.\\
 Section \ref{prelim} is dedicated to the reader less used with the tensorial formalism in handling multipolar expansions. For the informed reader, the section can be seen as an introduction to the notation and formulas used along the article.    In Sections \ref{estatic} and \ref{magnetost}  the results for the electric and magnetic fields  are presented in the static case. The dynamic case is treated  in Section \ref{dynamic}, and the  conclusions are outlined in  Section \ref{conclusion}. \\
 The formalism employed in this paper is a purely algebraic one. With a good understanding of the  definitions and notation presented in Section \ref{prelim}, we think the reader will be  able to verify every detail of the calculation  and, possibly, to search and find some more adequate versions.

\section{Preliminaries}\label{prelim}

\subsection{Definitions, notation and formulas from the tensorial formalism}
Let us consider a $n$-th order Cartesian tensor denoted by $\Tsft^{(n)}$ and characterized by the components 
$\Tsf_{i_1\dots i_n},\;\;i_q=1,2,3$. These components are, in fact, the components of a vector in the $n$ tensorial 
product of the Euclidean space $\mathbb{R}^3 $:
\beqan
\Tsft^{(n)}=\Tsf_{i_1\dots i_n}\,\evec_{i_1}\otimes\,\dots\,\otimes\evec_{i_n}=\Tsf_{i_1\dots i_n}\,
{\esft}_{i_1\dots i_n}\ ,
\eeqan
where $\evec_i$ are the unit vectors of a Cartesian basis in $\mathbb{R}^3$ and ${\esft}_{i_1\dots i_n} $ are the
unit vectors of the tensorial product space. Some vectors from the tensorial product of spaces 
can be represented by the tensorial products of $n$ vectors from the different space factors $\mathbb{R}^3$.  Particularly, in the case of  identical factors, we employ the usual notation
\beqan
\bbox{a}^n=a_{i_1}\dots a_{i_n}{\esft}_{i_1\dots i_n}\ .
\eeqan
The vector $\bbox{a}$ can be the differential operator $\nablav=\evec_i\,\d_i$ and, in this case,
\beqan
\left(\nablav^n\right)={\esft}_{i_1\dots i_n}\,\d_{i_1}\dots \d_{i_n}\ .
\eeqan
Because of this simplified notation, for avoiding confusions, the Laplace operator  in the space $\mathbb{R}^3 $ is symbolized  by $\Delta$, and not by the frequent  $\nablav^2$. \\
We employ the following notation for the tensorial contractions:
\beqan
 ({\Atens}^{(n)}||{\Btens}^{(m)})_{i_1 \cdots i_{|n-m|}}
=\left\{\begin{array}{ll}
A_{i_1 \cdots i_{n-m}j_1 \cdots j_m}B_{j_1 \cdots j_m} & ,\; n>m\\
A_{j_1 \cdots j_n}B_{j_1 \cdots j_n} & ,\; n=m\\
A_{j_1 \cdots j_n}B_{j_1 \cdots j_n i_1 \cdots i_{m-n}} & ,\; n<m
\end{array} \right.\ .
\eeqan
A fully symmetric tensor $\ssft^{(n)}$, which here is called simply "symmetric", has a projection on the subspace 
of symmetric and trace free ({\bf STF}) tensors. Up to a numerical factor, this projection will be represented by the tensor  $\Tcal(\ssft^{(n)})\equiv \bbox{\mathcal{S}}^{(n)}$. For  $n=2$, for example, we can write
\beqa\label{a.2}
\ssf_{ij}=\mathcal{S}_{ij}+\delta_{ij}\,\latens(\ssft^{(2)})\ .
\eeqa
The condition $\mathcal{S}_{ii}=0$ gives
 \beqa\label{a.3}
\lasf=\frac{1}{3}\ssf_{qq}\ .
\eeqa
In the case $n=3$:
\beqan
\ssf_{ijk}=\mathcal{S}_{ijk}+\delta_{\{ij}\lasf_{k\}}(\ssft^{(3)})\ ,
\eeqan
where  by the symbol $A_{\{i_1\dots i_n\}}$ is denoted the sum over all distinct permutations of the indexes. 
The condition $\mathcal{S}_{iik}=0$ for $k=1,2,3$ implies
 \beqan 
\lasf_i(\ssft^{(3)})=\frac{1}{5}\ssf_{qqi}\ .
 \eeqan

Though, maybe, only for a theoretical interest, let us consider  the general case for the {\bf STF} projection   of the symmetric tensor $\ssft^{(n)}$ defined up to a numerical factor by the equation
\beqan
\Tcal_{i_1\dots i_n}(\ssft^{(n)})\equiv \mathcal{S}_{i_1\dots i_n}=\ssf_{i_1\dots i_n}-\delta_{\{i_1i_2}\lasf_{i_3\dots i_n\}}(\ssft^{(n)})\ .
\eeqan
 $\latens^{(n-2)}\equiv\latens(\ssft^{(n)})$ is a symmetric tensor and is defined by the condition that $\Tcal^{(n)}$ is a trace-free 
tensor. For low values of $n$, the ones of really practical interest, the components $\lasf_{i_1\dots i_{n-2}}$ can be calculated directly from the equation system representing the vanishing relations of all the partial traces of the tensor $\Tcal^{(n)}$. However, we mention here a general formula known from literature \cite{Thorne,App} which, with 
the notation from the present paper, is written as
\beqan
&~&\fl\left[\Tcal\big[\ssft^{(n)}\big]\right]_{i_1\dots i_n}=
\suml^{[n/2]}_{m=0}\frac{(-1)^m(2n-1-2m)!!}{(2n-1)!!}\delta_{\{i_1i_2}\dots
\delta_{i_{2m-1}i_{2m}}\ssf^{(n:m)}_{i_{2m+1}\dots i_n\}} \ .
\eeqan
 The symbol $[\al]$ represents the integer part of $\al$ and $\ssf^{(n:m)}_{i_{2m+1}\dots i_n}$ denotes the components of the $(n-2\,m)$-th order tensor 
obtained from $\ssft^{(n)}$ by contracting $m$ pairs of symbols $i$. This equation is known as the {\it detracer theorem} \cite{App}. As a consequence of this theorem, the components of the tensor $\latens^{(n-2)}$ are written as
\beqan
\fl\lasf_{i_1\dots i_{n-2}}\big[\ssft^{(n)}\big]=
\suml^{[n/2-1]}_{m=0}\frac{(-1)^m[2n-1-2(m+1)]!!}{(m+1)(2n-1)!!}
\delta_{\{i_1i_2}\dots \delta_{i_{2m-1}i_{2m}}\ssf^{(n:\,m+1)}_{i_{2m+1}\dots i_{n-2}\}}.
\eeqan
These formulas are useful for defining and processing the multipole expansions of the electrodynamic potentials and fields.
\subsection{Multipole expansion of the electromagnetic field in Cartesian coordinates} 
The multipole expansions of the potentials in the Lorenz gauge are written as \cite{Castell,cv-sc,Gonzales,cv02}:
\beqa\label{a.9}
\Phi(\rvec,t)=\frac{1}{4\pi\eps_0}\suml_{n\ge0}\frac{(-1)^n}{n!}\,\nablav^{\,n}\vert\vert\frac{\psft^{(n)}(\tau)}{r}
\eeqa
and
\beqa\label{a.10}
\fl\;\;\;\;\;\;\;\;\;\;\;\;\Avec(\rvec,t)=\frac{\mu_0}{4\pi}\suml_{n\ge 1}\frac{(-1)^{n-1}}{n!}\left[\nablav\times\left(\nablav^{n-1}\vert\vert
\frac{\msft^{(n)}(\tau)}{r}\right)+\nablav^{n-1}\vert\vert\frac{\dot{\psft}^{(n)}(\tau)}{r}
\right]\ .
\eeqa
The dot symbolizes the time derivative and $\tau=t-r/c$ is the retarded time with respect  to the origin $O$ of the Cartesian axes in the point corresponding to the vector $\rvec$ . The origin $O$ is a point from the support of the electric charge and current  distribution. The tensors $\psft^{(n)}(t)$ and $\msft^{(n)}(t)$ are the electric and magnetic moments of the electric charges and currents 
distributions $\rho(\rvec,t)$ and $\jvec(\rvec,t)$, the support of these distributions being  included in the domain
 $\dom$:
 \beqan
\psft^{(n)}(t)=\int_{\dom}\rmd^3x\;\rvec^n\,\rho(\rvec,t),
  \eeqan  
  and
  \beqa\label{a.12}
\msft^{(n)}(t)=\frac{n}{n+1}\int_\dom \rmd^3x\;\rvec^n\times\jvec(\rvec,t)\ .
\eeqa
In the last equation,  a tensorial contraction via  the Levi-Civita pseudo-tensor $\eps_{ijk}$ is employed:
 \beqan
\left\{\Tsft^{(n)},\;\bbox{a}\right\}\,\to\, \Tsft^{(n)}\times \bbox{a}=\eps_{i_nqs}\Tsf_{i_1\dots i_{n-1}\,q}a_s
\esft_{i_1\dots i_n}\ ,
\eeqan
which in the particular case of $\Tsft^{(n)}=\bbox{b}^n$ becomes
\beqan
\bbox{b}^n\times\avec=b_{i_1}\,\dots\,b_{i_{n-1}}\left(\bbox{b}\times\avec\right)_{i_n}\esft_{i_1\dots i_n}\ .
\eeqan
The expansions \eref{a.9} and \eref{a.10} are running in the exterior of the minimal radius sphere including
 the support of $\rho$ and $\jvec$. \\
From equations \eref{a.9} and \eref{a.10} one obtains the following expansions of the fields $\Evec(\rvec,t)$ and $\Bvec(\rvec,t)$:
    \beqa\label{a.13} 
\fl\Evec(\rvec,t)&=&-\nablav\Phi(\rvec,t)-\frac{\d\Avec(\rvec,t)}{d t}=\frac{1}{4\pi\eps_0}\suml_{n\ge 0}
\frac{(-1)^{n-1}}{n!}\nablav^{n+1}\vert\vert\frac{\psft^{(n)}(\tau)}{r}\nonumber\\
\fl&&-\frac{\mu_0}{4\pi}\suml_{n\ge 1}
\frac{(-1)^{n-1}}{n!}\left[\nablav\times\left(\nablav^{n-1}\vert\vert\frac{\dot{\msft}^{(n)}(\tau)}{r}\right)
+\nablav^{n-1}\vert\vert\frac{\ddot{\psft}^{(n)}(\tau)}{r}\right]\,
\eeqa 
and
 \beqa\label{a.13a} 
\fl \Bvec(\rvec,t)=\nablav\times\Avec(\rvec,t)\nonumber\\
 \fl=\frac{\mu_0}{4\pi}\suml_{n\ge 1}\frac{(-1)^{n-1}}{n!}\left[\nablav^{n+1}\vert\vert
 \frac{\msft^{(n))}(\tau)}{r}-\nablav^{n-1}\vert\vert\Delta\frac{\msft^{(n)}(\tau)}{r}+\nablav\times\left(\nablav^{n-1}\vert\vert
 \frac{\dot{\psft}^{(n)}(\tau)}{r} \right) \right].
 \eeqa

For the magnetic moments \eref{a.12}, it is also possible to introduce {\bf STF} moments 
 $\mct^{(n)}=\Tcal(\msft^{(n)})$ but, this time, there are two steps required in order to complete the objective. The tensor $\msft^{(n)}$  is symmetric only in the first $n-1$ indexes    and satisfies the property
 \beqan   
\msf_{i_1\dots i_{n-2}\,qq}=0\ .
\eeqan
In the first step we must obtain the symmetric projection (up to a numerical factor) $\mlrt^{(n)}$ of the tensor
$\msft^{(n)}$. We begin with the first simple example corresponding to $n=2$.  Let us write the identity 
\beqan
\msf_{ij}=\frac{1}{2}\left(\msf_{ij}+ \msf_{ji}\right)+ \frac{1}{2}\left(\msf_{ij}- \msf_{ji}\right)=
\mlr_{ij}+\frac{1}{2}\eps_{ijk}\Nsf_k(\msft^{(2)})\ ,
\eeqan   
 where  $\mlrt^{(2)}$ is the symmetric part of $\msft^{(2)}$ and
 \beqan \Nsf_i(\msft^{(2)})=\eps_{ijk}M_{jk}=\frac{2}{3}\int_\dom\,\rmd^3x\,\left[\rvec\times\left(\rvec\times\jvec\right)\right]\ .
 \eeqan
 In this case ($n=2$), $\mct^{(2)}=\mlrt^{(2)}$ and, consequently, corresponds to the {\bf STF} projection. Therefore,
   \beqa\label{a.15}    
\msf_{ij}=\mc_{ij}+\frac{1}{2}\eps_{ijk}\Nsf_k(\msft^{(2)})\ .
 \eeqa
 For $n\ge 3$, we can generalize this result writing the identity:
 \beqa\label{a.16}  
\msf_{i_1\dots i_n}&=&\frac{1}{n}\left(\msf_{i_1\dots i_n}+\msf_{i_ni_2\dots i_{n-1} i_1}+\dots+ \msf_{i_1\dots i_n i_{n-1}}
\right)\nonumber\\
&+&\frac{1}{n}\left[\left(\msf_{i_1\dots i_n}-\msf_{i_n\dots i_{n-1}i_1}\right)+\dots\left(\msf_{i_1\dots i_n}
-\msf_{i_1\dots i_{n-2}i_n\,i_{n-1}}\right)\right]\nonumber\\
&=&\mlr_{i_1\dots i_n}+\frac{1}{n}\suml^{n-1}_{\la=1}\eps_{i_\la i_nq}
\Nsf^{(\la)}_{(i_1\dots i_{n-1}q)}(\msft^{(n)})\ ,
 \eeqa
 where  by $\Nsf^{(\la)}_{(i_1\dots i_{n-1}i_n)}$ we understand the component without the index $i_\la$  
 and the tensor $\Nsft^{(n-1)}=\Nsft(\msft^{(n)})$ is given by
\beqan    
\fl\;\;\;\;\;\;\Nsf_{i_1\dots i_{n-1}}(\msft^{(n)})=\eps_{i_{n-1}pq}\;\msf_{i_1\dots i_{n-2}pq}=\frac{n}{n+1}\int_{\dom}
\rmd^3x\,\,x_{i_1}\dots x_{i_{n-2}}\left[\rvec\times\left(\rvec\times\jvec\right)\right]_{i_{n-1}}\ .
 \eeqan
 It is a tensor of the same type as $\msft^{(n-1)}$,  i.e.\ symmetric in the first $n-2$ indices and with   
 $n-1$ vanishing  traces ($\Nsf_{i_1\dots i_{n-3}qq}=0$). Therefore, the {\bf STF} moment $\mct^{(n)}$ is given by 
 the components
\beqan    
\mc_{i_1\dots i_n}=\mlr_{i_1\dots i_n}-\delta_{\{i_1i_2}\lasf_{i_3\dots i_n\}}(\mlrt^{(n)})\ .
 \eeqan
As it will be seen in the dynamic case, we have to express  even the symmetric projection of the tensor $\Nsft(\msft^{(n)})$. For this,  it is useful to introduce the operator $\bbox{\mathcal N}$ defining the correspondence 
\beqan
\Nsft^{(n)}\,\to\,\bbox{\mathcal N}(\Nsft^{(n)}):\;\left(\bbox{\mathcal N}(\Nsft^{(n)})\right)_{i_1\dots i_{n-1}}
=\eps_{i_{n-1}ps}\;\Nsft_{i_1\dots i_{n-2}ps}\ .
\eeqan
Repeating this operation, we obtain:
\beqan    
\bbox{\mathcal N}^{2k}(\msft^{(n)})=\frac{(-1)^k\,n}{n+1}\int_{\dom}\rmd^3x\;r^{2k}\,\rvec^{n-2k}\times\jvec\ ,\nonumber\\
\bbox{\mathcal N}^{2k+1}(\msft^{(n)})=\frac{(-1)^k\,n}{n+1}\int_{\dom}\rmd^3x\;r^{2k}\,\rvec^{n-2k-1}\times(\rvec\times\jvec)\ .
\eeqan

\subsection{Delta-function identities and multipole  singularities}
 Usually, the multipole expansions are considered, term by term, as functions of $\rvec$ defined on $\mathbb{R}^3 $ 
 excepting a singular point which is chosen  as the origin $O$ of the Cartesian axes. Actually, these multipole terms are mathematical distributions (or generalized functions) considered firstly as regular ones having as support the entire space except the origin point $O$.  The observable quantities are expressed as weighted  averages on  spatial regions  or  as surface integrals of functions of field variables. No problems appear when in these regions  $r\ne 0$, but  when we have to calculate for example  the interaction of a distribution $(\rho,\,\jvec)$ with an external  field $(\Evec,\,\Bvec)$,
 \beqan
 W_{\scriptstyle int}=\int\rmd^3x\,\left(\rho\,\Phi-\jvec\cdot\Avec\right)
 \eeqan
 and the system associated with the external field $(\Evec,\,\Bvec)$ is represented by a point-like    multipole  system placed in $O$, the  singularities of the potentials or fields having as support this point become 
 unavoidable. As done in \cite{Frahm}, these singularities are determined starting from some $\delta$-function identities associated with the extension of multiple spatial derivatives of the functions of the type $1/r$ to the entire space. Having in mind the dynamic case, too,  we generalize such identities to the derivatives of the function $f(\tau)/r$. Representing the corresponding derivatives as functions of $\rvec$ and $t$ for $r\ne0$, we can write their  expressions in the form
\beqa\label{a.19} 
D_{i_1\dots i_n}(f)\equiv    
\d_{i_1}\dots\d_{i_n}\frac{f(\tau)}{r}=\suml^n_{l=0}\frac{1}{c^{n-l}r^{l+1}}C^{(n,\,l)}_{i_1\dots i_n}\,\frac{\rmd^{n-l}f(\tau)}{\rmd t^{n-l}}\ ,
\eeqa
where $C^{(n,\,l)}_{i_1\dots i_n}$ are fully symmetric in the indexes $i_1\dots i_n$. The general form of the coefficients $C$ is a simple consequence of the symmetry properties and of the derivative rules:
\beqa\label{a.20}     
C^{(n,\,l)}_{i_1\dots i_n}=\suml^{[\frac{n}{2}]}_{k=0}\,K^{(n,\,l)}_k\;\delta_{\{i_1i_2}\dots\delta_{i_{2k-1}i_{2k}}\,\nu_{2k+1}\dots \nu_{n\}}\ .
\eeqa
 For lower $n$ the coefficients $C$ and $K$ can be easily calculated by successive derivative operations. As one can  see in the following, for obtaining  general results for the $\delta$-type singularities of the multipole terms, it will be necessary to know only the coefficients 
\beqa\label{a.22}
K^{(n,\,n)}_0=(-1)^n\,(2n-1)!!\ .
 \eeqa
In \ref{A} we enumerate  the coefficients $C^{(n\,l)}_{i_1\dots i_n}$ for the first five values of $n$.\\
Considering the distribution representing the derivative \eref{a.19} extended to the entire space, we understand by 
 $\left(\d_{i_1}\dots\d_{i_n}\left(f(\tau)/r\right)\right)_{(0)}$ the singular part having as support the point $O$. In the following, we are interested  in writing explicitly  this contribution.\\
 A correct procedure for extracting this singular part is that employed in \cite{Frahm} defining 
\beqa\label{a.23}
\left\langle\,\left(\d_{i_1}\dots \d_{i_n}\frac{f(\tau)}{r}\right)_{(0)},\;\phi\right\rangle=\lim_{\eps\to 0}\int_{\dom_\eps}\rmd^3x
\d_{i_1}\dots \d_{i_n}\frac{f(\tau)}{r}\,\phi(\rvec) \nonumber\\
=\lim_{\eps\to 0}\left[\oint_{\Sigma_\eps}\rmd S\,\nu_{i_1}\d_{i_2}\dots \d_{i_n}\frac{f(\tau)}{r}
-\int_{\dom_\eps}\rmd^3x\,\d_{i_2}\dots \d_{i_n}\frac{f(\tau)}{r}\,\d_{i_1}\phi(\rvec)\right]\ .
 \eeqa 
 In this equation,  $\phi(\rvec)$ is supposed an element of the domain of distributions i.e.\ an  infinitely differentiable function. As it can be seen from equation \eref{a.19}, similar restrictive properties are considered for $f(\tau)$. The domain $\dom_\eps$ is the spherical region delimited  by the spherical surface $\Sigma_\eps$ with radius $\eps$. A basic hypothesis is the existence of the limits considered here, but the general demonstration of this property is a purely mathematical problem and bypasses the objectives   of the present paper.  Therefore,   the result of the limit considered is independent of the domains $\dom_\eps$ defined only by the condition $\dom_\eps\to O$.  
  Finally, for the calculation process in the present paper, the essential property is the possibility to represent the functions $f$ and $\phi$ in the domain $\dom_\eps$ by their Taylor series upon the origin $O$:
 \beqa\label{a.24}
\fl f(\tau)=\suml^\infty_{\la=0}\frac{(-1)^\la\, r^\la}{c^\la\la!}\left(\frac{\rmd^\la f(\tau)}{\rmd t^\la}\right)_{r=0},\nonumber\\
\fl\phi(\rvec)=\suml^\infty_{\al=0}\frac{1}{\al!}x_{i_1}\dots x_{i_\al}\left(\d_{i_1}\dots \d_{i_\al}\,\phi(\rvec)\right)
_{\rvec=0}=\suml^\infty_{\al=0}\frac{r^\al}{\al!}\nu_{i_1}\dots \nu_{i_\al}\left(\d_{i_1}\dots \d_{i_\al}\,\phi(\rvec)\right)_{\rvec=0}\ .
 \eeqa
 Equation \eref{a.23} can be written as
 \beqan
 \fl\left\langle\,\left(D_{i_1\dots i_n}(f)\right)_{(0)},\;\phi\right\rangle=\lim_{\eps\to 0}\oint_{\Sigma_\eps}\rmd S\,\nu_{i_1}\d_{i_2}\dots \d_{i_n}\frac{f(\tau)}{r}-\left\langle\,\left(D_{i_2\dots i_n}(f)\right)_{(0)},\;\d_{i_1}\phi\right\rangle
 \eeqan
 generating  a recursive calculation for a given $n$. \\
Let us express the $\delta$-singularities for lower order derivatives. Even in these simple cases, it becomes obvious the importance of  the angular average of a function $g(\nuvec)$ defined as 
\beqan
\langle g(\nuvec)\rangle=\frac{1}{4\pi}\int\,g(\nuvec)\,\rmd\Omega(\nuvec)\ .
\eeqan
For this average, we have the  well known    formula \cite{Thorne}:
\beqa\label{3.6}
\langle\nu_{i_1}\dots\nu_{i_n}\rangle=\left\{\begin{array}{c}0,
\;\;\;\;\;\;\;\;\;\;\;\;\;\;\;\;\;\;\;\;\;\;\;\;n=2k+1,\\
\frac{1}{(n+1)!!}\,\delta_{\{i_1i_2}\dots\delta_{i_{n-1}i_n\}},
\;\;\;\;\;\;\;\;\;\;n=2k,\;\;\;\;\;k=0,1,\dots\end{array}\right. 
\eeqa
The first $\delta$-singularity is obtaining for $n=2$ calculating the limit
\beqan
\fl\left\langle\,\left(D_{ij}(f)\right)_{(0)},\;\phi\right\rangle=\lim_{\eps\to0}\left[\oint_{\Sigma_\eps}\rmd S\,\nu_i\left(\d_j\frac{f(\tau)}{r}\right)\,\phi(\rvec)-\int_{\dom_\eps}\rmd^3x\,\left(\d_j\frac{f(\tau)}{r}\right)\,\d_i\phi(\rvec)\right]\\
\fl=\lim_{\eps\to0}\left[\oint_{\Sigma_\eps}\rmd S\,\nu_i\left(\d_j\frac{f(\tau)}{r}\right)\,\phi(\rvec)-\oint_{\Sigma_\eps}\frac{\rmd S}{r}\,\nu_j\,f(\tau)\,\d_i\phi(\rvec)+\int_{\dom_\eps}\frac{\rmd^3x}{r}\,f(\tau)\,\d_i\d_j\phi(\rvec)
\right]\ .
\eeqan
The first surface integral limit can be written as
\beqan
\bbox{L}_{1\sigma}=\lim_{\eps\to0}\oint\rmd\Omega(\nuvec)\;\eps^2\,\nu_i\left[-\frac{\nu_j}{c\eps} \dot{f}(\tau_\eps)
-\frac{\nu_j}{\eps^2}f(\tau_\eps)\right]\,\left[\phi(0)+\eps\nu_k\left(\d_k\phi\right)_0+\dots\right],
\eeqan
where $\tau_\eps=t-\eps/c$. Since $f(\tau_\eps)=f(t)+\mathcal{O}(\eps)$, and all the terms containing as factors positive powers of $\eps$ vanish at the limit $\eps\to0$, we can write
\beqan
\bbox{L}_{1\sigma}=-4\pi\left\langle\nu_i\nu_j\right\rangle\,f(t)\phi(0)=-\frac{4\pi}{3}\delta_{ij}\;f(t)\phi(0)\ .
\eeqan
The second surface integral limit 
\beqan
\bbox{L}_{2\sigma}=\lim_{\eps\to0}\oint\rmd\Omega(\nuvec)\,\eps\nu_j\,f(\tau_\eps)\left[\left(\d_i\phi\right)_0+\eps\nu_k\left(\d_i\d_k\phi\right)_0+\dots\right]\,=\,0\ ,
\eeqan
since all the terms contain as factors positive powers of $\eps$. As the  volume integral limit also vanishes,  we can  finally write:
 \beqa\label{3.5a}
\left(\d_i\d_j\frac{f(\tau)}{r}\right)_{(0)}=-\frac{4\pi}{3}f(t)\delta_{ij}\,\delta(\rvec)\ .
\eeqa
Including the expression of the derivative for $r\ne 0$, the above result written for $f(t)=1$ verifies equation \eref{1.3}.\\
Let us consider the next order derivative and calculate 
\beqan
\fl\left\langle\,\left(D_{ijk}(f)\right)_{(0)},\,\phi\right\rangle=\lim_{\eps\to0}\left[\oint_{\Sigma_\eps}\rmd S
\nu_i\left(\d_j\d_k\frac{f(\tau)}{r}\right)\,\phi(\rvec)-\int_{\dom_\eps}\rmd^3x\left(\d_j\d_k\frac{f(\tau)}{r}\right)\d_i\phi(\rvec)\right]=\\
\fl\lim_{\eps\to0}\left\{\oint\rmd\Omega(\nuvec)\,\eps^2\left[\frac{1}{c^2\eps}\nu_i\nu_j\nu_k\ddot{f}(\tau_\eps)+\frac{1}{c\eps^2}\nu_i(3\nu_j\nu_k-\delta_{jk})\dot{f}(\tau_\eps)
+\frac{1}{\eps^3}\nu_i(3\nu_j\nu_k-\delta_{jk})f(\tau_\eps)\right]\right.\\
\left. \times
\left(\phi(0)+\eps\nu_l\left(\d_l\phi\right)_0+\dots\right).
+\frac{4\pi}{3}\int_{\dom_\eps}\rmd^3x\,f(t)\delta_{jk}\delta(\rvec)\,\d_i\phi(\rvec)\right\}\ ,
\eeqan
where equation \eref{3.5a} is employed in the volume integral. In the surface integral limit, as $\eps\to 0$,  the   terms containing as factor the second order time derivative $\ddot{f}$ vanish because of the positive powers of $\eps$. The terms containing the first order time derivative $\dot{f}$ are proportional to  positive powers of $\eps$  for the term corresponding to $\phi(0)$, but in this case the corresponding limit cancels  since it contains also as factors the null averages $\langle\nu_i\nu_j\nu_k\rangle$ and $\langle\nu_i\rangle$. The only terms giving a non vanishing limit are provided by the product $f(\tau_\eps)\,\left(\d_l\phi\right)_0$ such that we can write 
\beqan
\fl\left\langle\,\left(D_{ijk}(f)\right)_{(0)},\,\phi\right\rangle &=&4\pi\left\langle3\nu_i\nu_j\nu_k\nu_l-\delta_{jk}\nu_i\nu_l\right\rangle\,f(t)\left(\d_l\phi\right)_0+\frac{4\pi}{3}f(t)\delta_{jk}\left(\d_i\phi\right)_0\\
&=&4\pi\left(\frac{1}{5}\delta_{\{ij}\delta_{kl\}}-\frac{1}{3}\delta_{jk}\delta_{il}\right)f(t)\left(\d_l\phi\right)_0
+\frac{4\pi}{3}f(t)\delta_{jk}\left(\d_l\phi\right)_0\ .
\eeqan
Finally:
\beqa\label{3.5b}
\left(\d_i\d_j\d_k\frac{f(\tau)}{r}\right)_{(0)}=-\frac{4\pi}{5}f(t)\delta_{\{ij}\d_{k\}}\delta(\rvec)\ .
\eeqa
For $f(t)=1$, equation \eref{3.5b} becomes the $\delta$-singularity corresponding to equation (13) from Ref.\ \cite{Frahm}. Equations \eref{3.5a}  and \eref{3.5b} will be applied in the static case for $f(t)=1$.

\section{Singularities of the electrostatic field}\label{estatic}
\par Let us consider the multipole expansion of the electrostatic field derived from the potential expansion \eref{a.9} in the static case:
 \beqan
\fl\;\;\;\;\;\Evec(\rvec)=\frac{1}{4\pi\eps_0}\suml_{n\ge 0}\frac{(-1)^{n-1}}{n!}\nablav^{n+1}\vert\vert\frac{\psft^{(n)}}{r}=
\frac{1}{4\pi\eps_0}\suml_{n\ge 0}\frac{(-1)^{n-1}}{n!}\psft^{(n)}\vert\vert\nablav^{n+1}\frac{1}{r}\ .
\eeqan
In the dipolar case ($n=1$), the potential has no $\delta$-singularities. For the electric field, we  apply equation \eref{3.5a}  obtaining the known result \eref{1.1}.\\
A first interesting feature of the $\delta$-singularities problem appears beginning with the electric quadrupole  potential and field. The singularity of the potential is obtained employing  equation \eref{3.5a}:
\beqan
\Phi^{(2)}(\rvec)=\left(\frac{1}{8\pi\eps_0}\nablav^2\vert\vert\frac{\psft^{(2)}}{r}\right)_{r\ne 0}-\frac{1}{6\eps_0}\psf_{qq}\delta(\rvec)\ .
\eeqan
The first term from the right-hand side of this equation is invariant to the substitution $\psft^{(2)}\,\to\,\pct^{(2)}=\Tcal(\psft^{(2)})$  and, employing equations \eref{a.2} and \eref{a.3} with $ \Ssft=\psft^{(2)}$, we can write
\beqan
\Phi^{(2)}(\rvec)
=\left(\frac{1}{8\pi\eps_0}\nablav^2\vert\vert\frac{\pct^{(2)}}{r}\right)_{r\ne0}
-\frac{1}{2\eps_0}\lasf\;\delta(\rvec)\ .
\eeqan
Usually, since  in the exterior of the charge distribution the potential $\Phi(\rvec)$ is solution of the Laplace equation, it is represented by a series of spherical functions:
\beqa\label{3.9}
\Phi(\rvec)=\suml^\infty_{l=0}\frac{1}{r^{l+1}}\suml^l_{m=-l}Q_{lm}\,Y_{lm}(\theta,\varphi)\ .
\eeqa 
The $2l+1$ spherical moments corresponding to a given $l$ are linear combinations of the $2l+1$ components of the $l$-th order {\bf STF} tensor $\pct^{(l)}$. For the above example of the 4-polar potential, it is obvious that using 
 the expansion representing the particular case of the equation \eref{3.9}, the $delta$-singularity of the potential is lost. The conclusion can be extended to the higher-order multipolar potentials. For the electrostatic field $\Evec(\rvec)$, the employment of the corresponding  expansion in spherical functions  leads to an incomplete description of the $\delta$-form singularities.\\
 Based on this observation, in our opinion, some care is necessary when passing to the representations by spherical functions since it is equivalent to working only in a subspace of the  mathematical objects used correctly in the theory. The  direct substitution $\psft\,\to\,\pct$ or passing to the spherical tensorial representations is not always justified. This    observation is pointed out  in another context in \cite{Raab}, too.\\
Searching the $\delta$-singularities of $\Evec^{(2)}(\rvec)$, instead of a straightforward processing of the corresponding expression in terms of the primitive moment $\psft^{(2)}$, we can introduce from the beginning the {\bf STF} moment $\pct^{(2)}$ writing
\beqan
\fl\;\;\;\;\;\;\;\;\;\left(\Evec^{(2)}(\rvec)\right)_{(0)}&=&-\frac{1}{8\pi\eps_0}\left(\nablav^3\vert\vert\frac{\psft^{(2)}}{r}
\right)_{(0)}=-\frac{1}{8\pi\eps_0}\left(\nablav^3\vert\vert\frac{\pct^{(2)}}{r}+\evec_i\d_i\Delta\frac{\lasf}{r}
\right)_{(0)}\nonumber\\
\fl\;\;\;\;\;\;\;\;\;&=&-\frac{1}{8\pi\eps_0}\left(\nablav^3\vert\vert\frac{\pct^{(2)}}{r}\right)_{(0)}+\frac{1}{2\eps_0}\lasf\delta(\rvec)\ ,
\eeqan
since $\Delta(1/r)=-4\pi\delta(\rvec)$. Equation \eref{3.5b}  finally gives
\beqa\label{3.10}
\fl\;\;\;\;\;\;\;\;\;\Evec^{(2)}(\rvec)=\left(\Evec^{(2)}(\rvec)\right)_{r\ne0}+\frac{1}{5\eps_0}\pct^{(2)}\vert\vert\nablav\delta(\rvec)+\frac{1}{2\eps_0}\lasf\nablav\delta(\rvec),\;\;\;\lasf=\latens(\psft^{(2)})\ ,
\eeqa
where 
\beqan
\left(\Evec^{(2)}(\rvec)\right)_{r\ne0}=-\frac{1}{8\pi\eps_0}\left(\nablav^3\vert\vert\frac{\psft^{(2)}}{r}\right)_{r\ne0}=-\frac{1}{8\pi\eps_0}\left(\nablav^3\vert\vert\frac{\pct^{(2)}}{r}\right)_{r\ne0}\ .
\eeqan
Even from this simple example it becomes obvious that this calculation version is convenient  for higher $n$ since some contractions of $\pct^{(n)}$ with the Kronecker symbols vanish. \\
In  case $n=3$, the advantage of inserting from the very beginning the {\bf STF} moments is more evident. Instead of searching the singularities of the fourth-order derivative $\d_i\d_j\d_k\d_l\,(1/r)$, we search the singularities of the contraction of this derivative with the {\bf STF} moment $\pct^{(3)}$ writing
\beqan
\fl\left\langle\left(\nablav^4\vert\vert\frac{\pct^{(3)}}{r}\right)_{(0)},\phi\right\rangle 
=\evec_i\lim_{\eps_0\to0}\left[\oint_{\Sigma_\eps}\rmd S\,\nu_i\d_j\d_k\d_l\frac{\pc_{jkl}}{r}\,\phi(\rvec)
-\int_{\dom_\eps}\rmd^3x\,\d_j\d_k\d_l\frac{\pc_{jkl}}{r}\,\d_i\phi(\rvec)\right].
\eeqan
For the surface integral limit, we apply equation \eref{a.19} particularized to the static case. Denoting by $\bbox{L}_\sigma$ this limit,
\beqan
\bbox{L}_{\sigma}=\evec_i\lim_{\eps\to 0}\oint_{\Sigma_\eps}\frac{\rmd\Omega(\nuvec)}{\eps^2}\nu_i\,C^{(3,\,3)}_{jkl}\pc_{jkl}\,\phi(\rvec)
\eeqan
and, inserting equation \eref{C3}, one obtains
\beqan
\bbox{L}_{\sigma}&=&\evec_i\lim_{\eps\to 0}\oint_{\Sigma_\eps}\frac{\rmd\Omega(\nuvec)}{\eps^2}\left(-15\nu_i\nu_j\nu_k\nu_l+3\nu_i\delta_{\{jk}\nu_{l\}}\right)\pc_{jkl}\,\phi(\rvec)\ .
\eeqan
Since $\delta_{\{jk}\nu_{l\}}\pc_{jkl}=0$, we get
\beqan
\fl\;\;\;\bbox{L}_{\sigma}&=&-15\evec_i\lim_{\eps\to 0}\oint_{\Sigma_\eps}\frac{\rmd\Omega(\nuvec)}{\eps^2}\nu_i\nu_j\nu_k\nu_l
\pc_{jkl}\left(\phi(0)+\eps\nu_q\left(\d_q\phi\right)_0+\frac{1}{2}\eps^2\nu_q\nu_s\left(\d_q\d_s\phi\right)_0\right)\\
\fl&=&-4\pi\times 15\,\evec_i\lim_{\eps\to 0}\left\langle\,\frac{1}{\eps^2}\nu_i\nu_j\nu_k\nu_l\,\pc_{jkl}
\left(\phi(0)+\eps\nu_q\left(\d_q\phi\right)_0+\frac{1}{2}\eps^2\nu_q\nu_s\left(\d_q\d_s\phi\right)_0\right)\right\rangle\ .
\eeqan
Denoting by $\al$ the power of $\eps$ in the Taylor series of $\phi(\rvec)$, we easily see  that for $\al\,\ge\,3$ all 
the terms from the last equation vanish for $\eps\,\to\,0$. For $\al=0$, the corresponding term vanishes since $\langle\nu_i\nu_j\nu_k\nu_l\rangle\,\pc_{jkl}=0$,  as it is seen by inserting equation \eref{3.6}. For $\al=1$, the related  term contains the average of an odd number of factors $\nu$ yielding a null result, too. The only non vanishing result corresponds to $\al=2$ such that 
\beqan
\bbox{L}_\sigma=-4\pi\times\frac{15}{2}\,\evec_i\left\langle\nu_i\nu_j\nu_k\nu_l\nu_q\nu_s\right\rangle\,\pc_{jkl}\ .
\eeqan
In the last equation we have the contraction  $\langle\nu_{i_1}\nu_{i_2}\nu_{i_3}\nu_{i_4}\nu_{i_5}\nu_{i_6}\rangle\pc_{i_4i_5i_6}$. It is not necessary to write explicitly the angular average of the six factors $\nu$ since, considering the general form of this  expression given in equation \eref{3.6}, we see that only the $3!=6$ terms of the form $\delta_{i_1i_j}\delta_{i_2i_k}\delta_{i_3i_l}$ with the  indexes $j,k,l$ covering  all the permutations of $(4,5,6)$ give non vanishing limits. Consequently,
\beqa\label{3.10a}
\bbox{L}_\sigma=-4\pi\frac{15\times3!}{2\times 7!!}\evec_i\pc_{ijk}\left(\d_j\d_k\phi\right)_0
=-\frac{24\pi}{14}\pct^{(3)}\vert\vert\left(\nablav^2\phi\right)_0\ .
\eeqa
The volume integral limit is zero since 
\beqan
\left(\nablav^3\vert\vert\frac{\pct^{(3)}}{r}\right)_{(0)}=0\ ,
\eeqan
as it can be easily verified. Equation \eref{3.10a} implies
\beqa\label{3.10bb}
\left(\nablav^4\vert\vert\frac{\pct^{(3)}}{r}\right)_{(0)}=-\frac{24}{14}\;\pct^{(3)}\vert\vert\nablav^2\delta(\rvec)\ .
\eeqa
 In  \ref{B} this equality is proven for the general case. 
Let us write 
\beqan
\fl\;\;\;\;\;\left(\Evec^{(3)}(\rvec)\right)_{(0)}=\frac{1}{24\pi\eps_0}\left(\nablav^4\vert\vert\frac{\psft^{(3)}}{r}\right)_{(0)}
=-\frac{1}{14\eps_0}\pct^{(3)}\vert\vert\nablav^2\delta(\rvec)-\frac{1}{2\eps_0}\bbox{\Lambda}\vert\vert\nabla^2\delta(\rvec)
\eeqan
or
\beqan
\Evec^{(3)}(\rvec)=\left(\Evec^{(3)}(\rvec)\right)_{r\ne 0}-\frac{1}{14\eps_0}\pct^{(3)}\vert\vert\nablav^2\delta(\rvec)-\frac{1}{2\eps_0}\bbox{\Lambda}\vert\vert\nabla^2\delta(\rvec)\ ,
\eeqan
where
\beqan
\left(\Evec^{(3)}(\rvec)\right)_{r\ne 0}=\frac{1}{24\pi\eps_0}\nablav^4\vert\vert\frac{\psft^{(3)}}{r}=
\frac{1}{24\pi\eps_0}\nablav^4\vert\vert\frac{\pct^{(3)}}{r}\ .
\eeqan
Even  if only for the  mathematical interest, we mention the possibility of expressing the multipole electrostatic field for arbitrary $n$. In \ref{B} it is introduced the general relation corresponding to equation \eref{3.10bb}:
\beqan
\left(\nablav^{n+1}\vert\vert\frac{\pct^{(n)}}{r}\right)_{(0)}=-\frac{4\pi\,n}{2n+1}\,\pct^{(n)}\vert\vert\nablav^{n-1}\delta(\rvec)
\eeqan
such that
\beqa\label{3.14}
\fl\;\;\;\;\;\;\left(\Evec^{(n)}(\rvec)\right)_{(0)}&=&\frac{(-1)^{n-1}}{4\pi\eps_0\,n!}\left(\nablav^{n+1}\vert\vert\frac{\pct^{(n)}}{r}+\frac{n(n-1)}{2}\nablav^{n-1}\vert\vert\Delta\frac{\latens^{(n-2)}}{r}\right)_{(0)}\nonumber\\
\fl\;\;\;\;\;\;&=&\frac{(-1)^n\,n}{\eps_0}\left(\frac{1}{n!\,(2n+1)}\pct^{(n)}+\frac{n-1}{2n!}\latens^{(n-2)}\right)\vert\vert\nablav^{n-1}\delta(\rvec)\ .
\eeqa
We point out that for $n\ge 4$,  for obtaining the complete  singular structure of the field,  we have to introduce  the irreducible tensors in the expression from equation \eref{3.14} corresponding to the $\delta$-singularities . 
This implies a recursive calculation introducing in equation \eref{3.14} the {\bf STF} projections of $\latens^{(n-2)},\,\latens^{(n-4)}, \dots$. For example, in the case $n=4$, we have to reduce the tensor $\latens^{(2)}=\latens(\psft^{(4)})$ writing:
\beqan
\latens^{(2)}\vert\vert\nablav^3\delta(\rvec)=\Tcal(\latens^{(2)})\vert\vert\nablav^3\delta(\rvec)+\latens(\latens^{(2)})\,\nablav\Delta\delta(\rvec)\ ,
\eeqan
where
\beqan
\lasf(\latens^{(2)})=\frac{1}{30}\psf_{qqss}\ ,\\
\mathcal{T}_{ij}(\latens^{(2)})=\frac{1}{7}\left(\psf_{qqij}-\frac{1}{3}\psf_{qqss}\delta_{ij}\right)\ .
\eeqan

\section{Singularities of the magnetostatic field}\label{magnetost}
\par Let us write the multipole expansions of the vector potential $\Avec(\rvec)$ and of the  magnetic field $\Bvec(\rvec)$ in the static case:
\beqan
\Avec(\rvec)&=&\frac{\mu_0}{4\pi}\suml_{n\ge 1}\frac{(-1)^{n-1}}{n!}\nablav\times\left(\nablav^{n-1}\vert\vert\msft^{(n)}\right)
\eeqan
and, respectively,
\beqa\label{4.3}
\Bvec(\rvec)=
\frac{\mu_0}{4\pi}\suml_{n\ge 1}\frac{(-1)^{n-1}}{n!}\left[\nablav\left(\nablav^n\vert\vert\frac{\msft^{n)}}{r}\right)
-\Delta\left(\nablav^{n-1}\vert\vert\frac{\msft^{(n)}}{r}\right)\right]\ .
 \eeqa
 For $r\ne 0$, the multipole magnetic field $\Bvec(\rvec)$ is expressed, as in the electric case, only by a scalar potential since the second term from the right-hand side of equation \eref{4.3} vanishes. However, the $\delta$-singularity of the magnetic field is not similar to the corresponding singularity of the electric field,   this second term giving the difference. The difference becomes easily obvious if one considers the fictitious magnetic shells (or sheets) employed in the Amp\`{e}re formalism. 
 It suffices to consider the case of the point-like magnetic dipole which can be taken as the limit of a current loop of infinitesimal size. For finite dimensions, the field of the loop  is derived from a scalar potential $\Phi_m$ 
 which is defined by an integral on the corresponding sheet and having a ``jump''  in all their points. Just this jump generates the  $\delta$-singularity corresponding to the second term from equation \eref{4.3}. An explicit calculation of this limit when the loop concentrates in a point is given in Ref.\ \cite{Corbo}.\\    
 In the dipolar case ($n=1$), retaining the singular term, we have
 \beqan
 \left(\Bvec^{(1)}(\rvec)\right)_{(0)}=\frac{\mu_0}{4\pi}\left(\evec_i\d_i\d_j\frac{m_j}{r}\right)_{(0)}+4\pi\,\mvec\delta(\rvec)\ .
 \eeqan
 Processing the first term as in the electrostatic case, we obtain  the well-known expression \eref{1.2}. 
No $\delta$-singularity is present in the dipolar vector potential.\\
Let us consider the quadrupole term from equation \eref{4.3} and the corresponding singularities:
\beqan
\left(\Bvec^{(2)}(\rvec)\right)_{(0)}=-\frac{\mu_0}{8\pi}\left(\nablav^3\vert\vert\frac{\msft^{(2)}}{r}-\nablav\vert\vert\left(\Delta\frac{\msft^{(2)}}{r}\right)\right)_{(0)}\ .
\eeqan
By introducing the {\bf STF} moment $\mc^{(2)}=\mlrt^{(2)}$ defined by equation \eref{a.15},
\beqan
\fl\left(\Bvec^{(2)}(\rvec)\right)_{(0)}&=&-\frac{\mu_0}{8\pi}\left(\nablav^3\vert\vert\frac{\mct^{(2)}}{r}
+\frac{1}{2}\evec_i\eps_{jkl}\d_i\d_j\d_k\frac{\Nsf_k}{r}-\evec_i\d_j\Delta\frac{\mc_{ji}}{r}
+\frac{1}{2}\eps_{ijl}\d_j\Delta\frac{\Nsf_l}{r}\right)_{(0)}\\
\fl &=&-\frac{3\mu_0}{10}\mct^{(2)}\vert\vert\nablav\delta(\rvec)-\frac{\mu_0}{4}\bbox{N}\times\nablav\delta(\rvec)\ ,
\eeqan
where $\bbox{N}=\evec_i\Nsf_i$. It follows that
\beqa\label{4.6}
\Bvec^{(2)}(\rvec)=\left(\Bvec^{(2)}(\rvec)\right)_{r\ne0}
-\frac{3\mu_0}{10}\,\mct^{(2)}\vert\vert\nablav\delta(\rvec)-\frac{\mu_0}{4}\,\bbox{N}\times\nablav\delta(\rvec)
\eeqa
where 
\beqan
\left(\Bvec^{(2)}(\rvec)\right)_{r\ne0}=\left(-\frac{\mu_0}{8\pi}\nablav^3\vert\vert\frac{\mct^{(2)}}{r}\right)_{r\ne0}=\left(-\frac{\mu_0}{8\pi}\nablav^3\vert\vert\frac{\msft^{(2)}}{r}\right)_{r\ne0}\ .
\eeqan
 
Let us consider the quadrupolar vector potential $\Avec^{(2)}(\rvec)$ with equation \eref{a.15} already inserted:
 \beqa\label{4.8} 
\Avec^{(2)}(\rvec)&=&-\frac{\mu_0}{8\pi}\evec_i\eps_{ijk}\,\d_j\d_l\frac{\msf_{lk}}{r}
=-\frac{\mu_0}{8\pi}\nablav\times\left(\nablav\vert\vert\frac{\mct^{(2)}}{r}\right)\nonumber\\
&&+\frac{\mu_0}{16\pi}\nablav\left(\nablav\cdot\frac{\bbox{N}}{r}\right)
-\frac{\mu_0}{16\pi}\bbox{N}\Delta\frac{1}{r}\ .
  \eeqa
 The term proportional to $\Delta(1/r)$ vanishes   for $r\ne 0$. Whereas $\Bvec^{(2)}$ is invariant to the substitution $\psft^{(2)}\to \pct^{(2)}$, this property is not verified by the vector potential $\Avec^{(2)}$. 
 This substitution has as result an additional gradient term, i.e.\ a gauge transformation of the potential. 
Though, by the extension to the entire space, the gradient term  from expression \eref{4.8} generates  a $\delta$-singularity, this term gives no contribution to the singularities of the magnetic field The singularities of $\Bvec$  are, by definition, only the singularities corresponding to the curl of the vector potential. It remains still the problem of  independence 
of the physical results on the $\Avec$ $\delta$-singularities corresponding to different gauges. Maybe, the presence of such singularities in the expression of an interaction Hamiltonian corresponding to the density term $ \jvec\cdot\Avec$ can generate the problem of proving the gauge independence of physical results as in the case of the Aharonov-Bohm effect.\\
We can give, as in the electrostatic case, the general formula for the singularities of the magnetic field writing
 \beqa\label{4.10} 
\left(\Bvec^{(n)}(\rvec\right)_{(0)}=\frac{(-1)^{n-1}\mu_0}{4\pi \,n!}\left(\nablav^{n+1}\vert\vert\frac{\msft^{(n)}}{r}-
\nablav^{n-1}\vert\vert\Delta\frac{\msft^{(n)}}{r}\right)_{(0)}\ .
\eeqa
The introduction of the symmetric moment $\mlrt^{(n)}$, equation \eref{a.16}, gives
\beqan
\fl\;\;\;\;\;\;\nablav^{n+1}\vert\vert\frac{\msft^{(n)}}{r}=
\nablav^{n+1}\vert\vert\frac{\mlrt^{(n)}}{r}
+\frac{1}{n}\evec_i\d_i\d_{i_1}\dots\d_{i_n}\suml^{n-1}_{\la=1}\eps_{i_\la i_n q}
\frac{\Nsf^{(\la)}_{(i_1\dots  i_{n-1}q)}}{r}=\nablav^{n+1}\vert\vert\frac{\mlrt^{(n)}}{r}\ ,
\eeqan
since in all terms of the sum there are the contractions $\eps_{i_li_nq}\d_{i_l}\d_{i_n}$ for $l=1,\dots n-1$ which vanish. From equation \eref{3.14} we can write a corresponding result for the magnetic field with the substitutions $\psft^{(n)}\to \mlrt^{(n)},\;\pct^{(n)}\to\mct^{(n)},\;\eps_0\to 1/\mu_0$:
\beqan 
\fl\;\;\;\;\;\;\frac{(-1)^{n-1}\mu_0}{4\pi\,n!}\left(\nablav^{n+1}\vert\vert\frac{\mlrt^{(n)}}{r}\right)_{(0)}&=&
(-1)^n\mu_0\left(\frac{n}{n!\,(2n+1)}\mct^{(n)}\vert\vert\nablav^{n-1}\delta(\rvec)\right.\nonumber\\
\fl\;\;\;\;\;\;&&+\left.\frac{n(n-1)}{2\,n!}\widetilde{\latens}^{(n-2)}\vert\vert\nablav^{n-1}\delta(\rvec)\right)\ ,
\eeqan
where $\widetilde{\latens}^{(n-2)}=\latens(\mlrt^{(n)})$. 
The second term from equation \eref{4.10} is processed as follows:
\beqan
\nablav^{n-1}\vert\vert\Delta\frac{\msft^{(n)}}{r}&=\nablav^{n-1}\vert\vert\Delta\frac{\mlrt^{(n)}}{r}
+\frac{1}{n}\evec_i\d_{i_1}\dots\d_{i_{n-1}}\suml^{n-1}_{\la=1}\eps_{i_\la iq}\Delta\frac{\Nsf^{(\la)}_{(i_1\dots i_{n-1}q)}}{r}\\
&=\nablav^{n-1}\vert\vert\Delta\frac{\mlrt^{(n)}}{r}+\frac{n-1}{n}\evec_i\,\eps_{iqi_{n-1}}\,\Nsf_{i_1\dots i_{n-2}q}\d_{i_1}\dots\d_{i_{n-1}}\Delta\frac{1}{r}\ .
\eeqan
 The term containing $\mlrt$ can be further transformed to:
\beqan
\fl&&\mlrt^{(n)}\vert\vert\nablav^{n-1}\delta(\rvec)=\mct^{(n)}\vert\vert\nablav^{n-1}\delta(\rvec)+
\evec_i\,\d_{i_1}\dots\d_{i_{n-1}}\,\delta(\rvec)\delta_{\{i_1i_2}\widetilde{\lasf}_{i_3\dots i_{n-1}\,i\}}\\
\fl&=&\mct^{(n)}\vert\vert\nablav^{n-1}\delta(\rvec)+\frac{(n-1)(n-2)}{2}\evec_i\d_{i_1}\dots\d_{i_{n-3}}\Delta\delta(\rvec)\widetilde{\lasf}_{i_1\dots i_{n-3}i}\\
\fl&&+(n-1)\evec_i\d_i\,\d_{i_1}\dots \d_{i_{n-2}}\delta(\rvec)\widetilde{\lasf}_{i_1\dots i_{n-2}}\\
\fl&=&\mct^{(n)}\vert\vert\nablav^{n-1}\delta(\rvec)+\frac{(n-1)(n-2)}{2}\widetilde{\latens}^{n-2}\vert\vert\nablav^{n-3}\Delta\delta(\rvec)+(n-1)\widetilde{\latens}^{n-2}\vert\vert\nablav^{n-1}\delta(\rvec)\ .
\eeqan
Finally,
\beqa\label{finmst}
\fl\;\;\;\;\;\left(\Bvec^{(n)}(\rvec)\right)_{(0)}&=&\frac{(-1)^n\mu_0}{n!}\left[\left(-\frac{n+1}{2n+1}\mct^{(n)}+\frac{(n-1)(n-2)}{2}\widetilde{\latens}^{(n-2)}\right)\vert\vert\nablav^{n-1}\delta(\rvec)\right.\nonumber\\
\fl\;\;\;\;\;&-&\left.\frac{n-1}{n}\Nsft^{(n-1)}\vert\times\vert\nablav^{n-1}\delta(\rvec)
-\frac{(n-1)(n-2)}{2}\widetilde{\latens}^{(n-2)}\vert\vert\nablav^{n-3}\Delta\delta(\rvec)
\right]
\eeqa
where  the notation 
\beqan
\left(\Asft^{(n)}\vert\times\vert\Bsft^{(n)} \right)_{i_1\dots i_n}=\eps_{i_nqs}\Asf_{i_1\dots i_{n-1}\,q}\Bsf_{i_1\dots i_{n-1}\,s}
\eeqan
is introduced. The general expression is written  as
\beqan
\Bvec^{(n)}(\rvec)=\left(\Bvec^{(n)}(\rvec)\right)_{r\ne 0}+\left(\Bvec^{(n)}(\rvec)\right)_{(0)}(\rvec)\ ,
\eeqan
where
\beqan
\left(\Bvec^{(n)}(\rvec)\right)_{r\ne 0}=\frac{(-1)^{n-1}\mu_0}{4\pi\,n!}\nablav^{n+1}\vert\vert\frac{\mct^{(n)}}{r}
=\frac{(-1)^{n-1}\mu_0}{4\pi\,n!}\nablav^{n+1}\vert\vert\frac{\msft^{(n)}}{r}\ ,
\eeqan
with the derivatives calculated for $r\ne 0$.\\
Here is the place for the  same observation as at the end of Section \eref{estatic}, but in the singular term containing $\Nsft^{(n-1)}$
the reduction to the {\bf STF} tensor $\Tsft^{(n-1)}$ begins from $n=3$. Particularly,  
\beqan
 \Tcal(\Nsft^{(2)})=\stackrel{<->}{\Nsft}^{(2)},\;\;\;  \stackrel{<->}{\Nsf}_{ij}=\frac{1}{2}\left(\Nsf_{ij}+\Nsf_{ji}\right)\ ,
\eeqan
and
\beqan
\fl\;\;\;\;\;\;\Nsft^{(2)}\vert\times\vert\nablav^2\delta(\rvec)=\stackrel{<->}{\Nsft}^{(2)}\vert\times\vert\nablav^2\delta(\rvec)
+\frac{1}{2}\Nmc(\Nsft^{(2)})\vert\vert\nablav^2\delta(\rvec)-\frac{1}{2}\Nmc(\Nsft^{(2)})\Delta\delta(\rvec)\ .
\eeqan
Obviously, the analysis of the  higher order singular terms    is more complicated for the magnetic field  than for the  electric one,  but it is  purely a technical problem.

\section{The dynamic case}\label{dynamic}
Considering the $n$-th order multipole  electric and magnetic fields, writing separately the  contributions of  the electric and magnetic multipoles,

 \beqa\label{5.1} 
 \Evec^{(n,\,p)}(\rvec,t)=\frac{(-1)^{n-1}}{4\pi\eps_0\,n!}\left(\nablav^{n+1}\vert\vert\frac{\psft^{(n)}(\tau)}{r}
-\frac{1}{c^2}\nablav^{n-1}\vert\vert\frac{\ddot{\psft}^{(n)}(\tau)}{r}\right),\nonumber\\
\Evec^{(n,\,m)}(\rvec,t)=\frac{(-1)^n}{4\pi\eps_0c^2\,n!}\nablav\times\left(\nablav^{n-1}\vert\vert\frac{\dot{\msft}^{(n)}(\tau)}{r}\right)
\eeqa
 and
  \beqa\label{5.1a}
 \Bvec^{(n,\,p)}(\rvec,t)=\frac{(-1)^{n-1}\mu_0}{4\pi\,n!}\nablav\times\left(\nablav^{n-1}\vert\vert\frac{\dot{\psft}^{(n)}(\tau}{r}\right)\ ,\nonumber\\ \Bvec^{(n,\,m)}(\rvec,t)=\frac{(-1)^{n-1}\mu_0}{4\pi\,n!}\left(\nablav^{n+1}\vert\vert\frac{\msft^{(n)}(\tau)}{r}-\nablav^{n-1}\vert\vert\Delta\frac{\msft^{(n)}(\tau)}{r}
\right)\ ,
 \eeqa
we search the corresponding $\delta$-singularities. Processing  the different   terms from equation \eref{5.1} and \eref{5.1a} for reducing the moment tensors to the {\bf STF} ones, we deal with the Laplace operator $\Delta$ applied to functions of the type $f(\tau)/r$. Since the equation verified by these functions is
 \beqa\label{5.2} 
\Delta\frac{f(\tau)}{r}-\frac{1}{c^2}\frac{\d^2}{\d t^2}\frac{f(\tau)}{r}=-4\pi f(t)\delta(\rvec)\ ,
\eeqa
the processing is more complicated in the dynamic case compared with the static one. \\
 For the electric field of the electric dipole we write the singular part as
 \beqan
\fl\left(\Evec^{(1,\,p)}(\rvec,t)\right)_{(0)}=\frac{1}{4\pi\eps_0}\left[\nablav\left(\nablav\cdot\frac{\pvec(\tau)}{r}\right)-\frac{1}{c^2}\frac{\ddot{\pvec}(\tau)}{r}\right]_{(0)}=\frac{1}{4\pi\eps_0}\,\evec_i\,\left(\d_i\d_j\frac{p_j(\tau)}{r}\right)_{(0)}\ .
  \eeqan
Note that  the second term proportional to $\ddot{\pvec}$ has no $\delta$-singularity. Equation \eref{3.5a} gives a result similar to that from the static case:
 \beqan 
 \left(\Evec^{(1,\,p)}(\rvec,t)\right)_{(0)}=-\frac{1}{3\eps_0}\,\pvec(t)\,\delta(\rvec)\ .
\eeqan 
The magnetic field $\Bvec^{(1,\,p)}(\rvec,t)$ of the electric dipole has no $\delta$-singularity.\\
Obviously, the electric field $\Evec^{(1,\,m)}(\rvec,t)$ has no $\delta$-singularity either. Writing the singular part of $\Bvec^{(1,\,m)}(\rvec,t)$,
\beqan
\left(\Bvec^{(1,\,m)}(\rvec,t)\right)_{(0)}=\frac{\mu_0}{4\pi}\left[\nablav\left(\nablav\cdot\frac{\mvec(\tau)}{r}\right)-\frac{1}{c^2}\frac{\ddot{\mvec}(\tau)}{r}+4\pi\,\mvec(t)\delta(\rvec)
\right]_{(0)}\ ,
\eeqan
one obtains the result similar to that from the static case:
 \beqan 
 \left(\Bvec^{(1,\,m)}(\rvec,t)\right)_{(0)}=\frac{2\mu_0}{3}\mvec(t)\,\delta(\rvec)\ .
 \eeqan
Let us consider the electric field $\Evec^{(2,\,p)}(\rvec,t)$ and search the corresponding $\delta$-singularities by  introducing firstly the {\bf STF} moment $\pct^{(2)}$:
\beqan
\fl\left(\Evec^{(2,\,p)}(\rvec,t)\right)_{(0)}=-\frac{1}{8\pi\eps_0}\left[\nablav^3\vert\vert\frac{\psft^{(2)}(\tau)}{r}-\frac{1}{c^2}\nablav\vert\vert\frac{\ddot{\psft}^{(2)}(\tau)}{r}\right]_{(0)}\\
\fl\;\;\;\;\;\;\;\;\;\;\;\;\;\;\;\;=-\frac{1}{8\pi\eps_0}\left[\nablav^3\vert\vert\frac{\pct^{(2)}(\tau)}{r}-\frac{1}{c^2}\nablav\vert\vert\frac{\ddot{\pct}^{(2)}(\tau)}{r}+\nablav\left(\Delta\frac{\lasf(\tau)}{r}\right)-\frac{1}{c^2}\nablav\frac{\ddot{\lasf}(\tau)}{r}\right]_{(0)}\ .
\eeqan
The insertion of equation \eref{5.2} gives
\beqan
\Evec^{(2,\,p)}_{(0)}=-\frac{1}{8\pi\eps_0}\left[\nablav^3\vert\vert\frac{\pct^{(2)}(\tau)}{r}-\frac{1}{c^2}\nablav\vert\vert\frac{\ddot{\pct}^{(2)}(\tau)}{r}-4\pi\,\lasf(t)\nablav\delta(\rvec)\right]_{(0)}\ .
\eeqan
Expressing the singularity of the first term from the parenthesis by the insertion of equation \eref{3.5b}, we can finally write 
  \beqa\label{5.4} 
\Evec^{(2,\,p)}(\rvec,t)=\left(\Evec^{(2,\,p)}(\rvec,t)\right)_{r\ne0}+\frac{1}{5\eps_0}\pct^{(2)}(t)\vert\vert\nablav\delta(\rvec)+\frac{1}{2\eps_0}\lasf(t)\delta(\rvec)\ ,
\eeqa
where
  \beqan
\left(\Evec^{(2,\,p)}(\rvec,t)\right)_{r\ne0}=-\frac{1}{8\pi\eps_0}\left[\nablav^3\vert\vert\frac{\pct^{(2)}(\tau)}{r}-\frac{1}{c^2}\nablav\vert\vert\frac{\ddot{\pct}^{(2)}(\tau)}{r}\right]_{r\ne0}\ .
\eeqan
The result \eref{5.4} is similar to the result \eref{3.10} from the static case. For the electric 4-polar term, the electric field expression for $r\ne 0$ is invariant to the substitution $\psft^{(2)}\,\to\,\pct^{(2)}$ as in the static case but, as we will see in the following, such property  is not yet  verified for higher order multipoles in the dynamical case.\\
Searching the singularities corresponding to the electric field $\Evec^{(2,\,m)}$ of the magnetic quadrupole, the difference from the static case is obvious. Let us express this field with the help of  the {\bf STF} magnetic moment $\mct^{(2)}=\mlrt^{(2)}$:
\beqan
\fl\Evec^{(2,\,m)}(\rvec,t)=\frac{1}{8\pi\eps_0c^2}\,\nablav\times\left(\nablav\vert\vert\frac{\dot{\msft}^{(2)}}{r}\right)=\frac{1}{8\pi\eps_0c^2}\left[\nablav\times\left(\nablav\vert\vert\frac{\dot{\mct}^{(2)}}{r}\right)\right.\\
\fl+\left.\frac{1}{2}\evec_i\eps_{ijk}\eps_{lkq}\,\d_j\d_l\frac{\dot{\Nsf}_q(\tau)}{r}
\right]=\frac{1}{8\pi\eps_0c^2}\left[\nablav\times\left(\nablav\vert\vert\frac{\dot{\mct}^{(2)}}{r}\right)
-\frac{1}{2}\nablav^2\vert\vert\frac{\dot{\bbox{N}}(\tau)}{r}+\frac{1}{2}\Delta\frac{\dot{\bbox{N}}(\tau)}{r}
\right]\ .
\eeqan 
We employed the same notation as in equation \eref{4.6}. The singularities added by the extension of this expression to the entire space are the following:
 \beqa\label{5.6} 
\fl\;\;\;\;\;\;\;\left[\nablav\times\left(\nablav\vert\vert\frac{\dot{\mct}^{(2)}}{r}\right)\right]_{(0)}=
\evec_i\eps_{ijk}\left[\d_j\d_l\frac{\dot{\mc}_{lk}}{r}\right]_{(0)}=-\frac{4\pi}{3}\evec_i\eps_{ijk}\dot{\mc}_{jk}(t)\delta(\rvec)=0\ ,\nonumber\\
\fl\;\;\;\;\;\;\;\;\;\;\;\;\;\;\;\;\;\;\;\left[\nablav^2\vert\vert\frac{\dot{\bbox{N}}(\tau)}{r}\right]_{(0)}=-\frac{4\pi}{3}\dot{\bbox{N}}(t)\,\delta(\rvec)\ .
\eeqa
For  the last term from the expression of $\Evec^{(2,\,m)}$, we have to consider the equation \eref{5.2}, i.e.\
\beqan
\Delta\frac{\dot{\bbox{N}}(\tau)}{r}=\frac{1}{c^2}\frac{\tdot{\bbox{N}}(\tau)}{r}-4\pi\dot{\bbox{N}}(t)\,\delta(\rvec)\ ,
\eeqan
such that we can write
\beqa\label{5.7} 
\fl\Evec^{(2,\,m)}(\rvec,t)=\frac{1}{4\pi\eps_0} \left[\frac{1}{2c^2}\nablav\times\left(\nablav\vert\vert\frac{\dot{\mct^{(2)}}(\tau)}{r}\right) - \left(\frac{1}{4c^2}\nablav^2\vert\vert\frac{\dot{\bbox{N}}(\tau)}{r}-\frac{1}{4c^4}\frac{\tdot{\bbox{N}}(\tau)}{r}\right)\right]_{r\ne0} \nonumber \\
+ \frac{1}{12\eps_0c^2}\dot{\bbox{N}}(t)\delta(\rvec)-\frac{1}{4\eps_0 c^2}\dot{\bbox{N}}(t)\delta(\rvec)\ .
\eeqa
In the above  result one can notice that for $r\ne0$, the expression of the electric field is not invariant to the substitution $\psft^{(2)}\,\to\,\pct^{(2)}$. The additional  term which appears as a consequence of this substitution is exactly the expression of the electric field corresponding to an electric dipole with the moment
 \beqa\label{5.8} 
\delta\pvec^{\,'}=-\frac{1}{4c^2}\dot{\bbox{N}}\ .
\eeqa
In equation \eref{5.7}, we have written separately the term  which represents the $\delta$-singularity corresponding to this dipole. \\
Let us express the field $\Bvec^{(2)}(\rvec,t)$. For the part $\Bvec^{(2,\,p)}$, we insert the {\bf STF} moment $\pct^{(2)}$ which gives
 \beqa\label{5.9b} 
\fl\Bvec^{(2,\,p)}(\rvec,t)&=&\frac{\mu_0}{8\pi}\nablav\times\left(\nablav\vert\vert\frac{\dot{\psft}^{(2)}(\tau)}{r}\right)
=\frac{\mu_0}{8\pi}\left[\nablav\times\left(\nablav\vert\vert\frac{\dot{\pct}^{(2)}(\tau)}{r}\right)+\evec_i\eps_{ijk}\d_j\d_l\frac{\lasf(\tau)}{r}\delta_{lk}\right]\nonumber\\
\fl&=&\frac{\mu_0}{8\pi}\nablav\times\left(\nablav\vert\vert\frac{\dot{\pct}^{(2)}(\tau)}{r}\right)\ ,
\eeqa
since $\eps_{ijk}\d_j\d_l\delta_{lk}=\eps_{ijk}\d_j\d_k=0$. It follows that $\Bvec^{(2,\,p)}$ is invariant to the substitution $\psft^{(2)}\,\to\,\pct^{(2)}$ and has no $\delta$-singularities due to equation \eref{5.6} written for  $\mct^{(2)}\,\to\,\pct^{(2)}$.\\
The introduction of the {\bf STF} magnetic moment $\mct^{(2)}=\mlrt^{(2)}$ in the expression of $\Bvec^{(2,\,m)}$ results in
\beqan
\fl\Bvec^{(2,\,m)}(\rvec,t)&=&-\frac{\mu_0}{8\pi}\left(\nablav^3\vert\vert\frac{\msft^{(2)}(\tau)}{r}-\nablav\vert\vert\Delta\frac{\msft^{(2)}(\tau)}{r}\right)\\
\fl&=&-\frac{\mu_0}{8\pi}\left(\nablav^3\vert\vert\frac{\mct^{(2)}(\tau)}{r}
-\nablav\vert\vert\Delta\frac{\mct^{(2)}(\tau)}{r}+\frac{1}{2}\nablav\times\Delta\frac{\bbox{N}(\tau)}{r}\right)\ .
\eeqan
Separating the $\delta$-singularities with the help of  equations \eref{3.5b} and \eref{5.2}, we obtain
 \beqa\label{5.9} 
\fl\Bvec^{(2,\,m)}(\rvec,t)&=&-\frac{\mu_0}{8\pi}\left(\nablav^3\vert\vert\frac{\mct^{(2)}(\tau)}{r}-\frac{1}{c^2}\nablav\vert\vert\frac{\ddot{\mct}^{(2)}(\tau)}{r}+\frac{1}{2c^2}\nablav\times\frac{\ddot{\bbox{N}}(\tau)}{r}\right)_{r\ne 0}\nonumber\\
\fl&&-\frac{3\mu_0}{10}\mct^{(2)}(t)\vert\vert\nablav\delta(\rvec)-\frac{\mu_0}{4}\bbox{N}(t)\times\nablav\,\delta(\rvec)\ .
\eeqa
Since for $r\ne0$ the magnetic field of the magnetic quadrupole  can be written as
\beqan
\left(\Bvec^{(2,\,m)}(\rvec,t)\right)_{r\ne0}=-\frac{\mu_0}{8\pi}\left(\nablav^3\vert\vert\frac{\msft^{(2)}(\tau)}{r}-\frac{1}{c^2}\nablav\vert\vert\frac{\ddot{\msft}^{(2)}(\tau)}{r}\right)_{r\ne0}\ ,
\eeqan
we can conclude that for $r\ne0$ this field is not invariant to the substitution $\msft^{(2)}\,\to\,\mct^{(2)}$ and the additional term introduced by this substitution is equivalent to that of an electric dipole having the moment $\delta\pvec^{\,'}$ given by equation \eref{5.8}. Obviously, the corresponding term has no $\delta$-singularity such that it is not represented by a singular term in equation \eref{5.9}.\\
Let us go further to the field of  the third order electric multipole. Beginning with the electric field, we have to extend to the entire space the expression 
 \beqa\label{5.9a} 
\Evec^{(3,\,p)}(\rvec,t)=\frac{1}{24\pi\eps_0}\left(\nablav^4\vert\vert\frac{\psft^{(3)}(\tau)}{r}
-\frac{1}{c^2}\nablav^2\vert\vert\frac{\ddot{\psft}^{(3)}(\tau)}{r}\right)\ .
\eeqa
We consider separately each term from the last equation introducing the {\bf STF} moment $\pct^{(3)}$:
\beqan
\fl\nablav^4\vert\vert\frac{\psft^{(3)}(\tau)}{r}=\nablav^4\vert\vert\frac{\pct^{(3)}(\tau)}{r}
+\evec_i\d_i\,\d_{i_1}\d_{i_2}\d_{i_3}\frac{\delta_{\{i_1i_2}\lasf_{k\}}(\tau)}{r}
=\nablav^4\vert\vert\frac{\pct^{(3)}(\tau)}{r}+3\evec_i\d_i\d_j\frac{\lasf_j(\tau)}{r}\ .
\eeqan
Inserting equation \eref{D9} for the $\delta$-singularity of the first term from the right-hand side of the last equation and employing equation \eref{5.2} for the second one, we can write  
 \beqa\label{5.10} 
\fl\;\;\;\;\;\;\;\;\;\nablav^4\vert\vert\frac{\psft^{(3)}(\tau)}{r}&=&\left(\nablav^4\vert\vert\frac{\pct^{(3)}(\tau)}{r}
+\frac{3}{c^2}\nablav^2\vert\vert\frac{\ddot{\bbox{\Lambda}}(\tau)}{r} \right)_{r\ne 0}\nonumber\\
\fl\;\;\;\;\;\;\;\;\;&&-\frac{12\pi}{7}\pct^{(3)}(t)\vert\vert\nablav^2\delta(\rvec)-\frac{4\pi}{c^2}\ddot{\bbox{\Lambda}}(t)\delta(\rvec)
-12\pi\bbox{\Lambda}(t)\vert\vert\nablav^2\delta(\rvec)\ ,
\eeqa
where $\bbox{\Lambda}=\lasf_i\evec_i$. Regarding the second term from equation \eref{5.9a}, we similarly obtain 
 \beqa\label{5.11} 
\fl\;\;\;\;\;\;\; \nablav^2\vert\vert\frac{\ddot{\psft}^{(3)}(\tau)}{r}=\left(\nablav^2\vert\vert\frac{\ddot{\pct}^{(3)}(\tau)}{r}
 +2\,\nablav^2\vert\vert\frac{\ddot{\bbox{\Lambda}}(\tau)}{r}+\frac{1}{c^2}\frac{\qdot{\bbox{\Lambda}}(\tau)}{r}
 \right)_{r\ne 0}-\frac{20\pi}{3}\ddot{\bbox{\Lambda}}(t)\delta(\rvec)
 \eeqa
since the first term containing $\nablav^2\vert\vert(\pct^{(3)}(\tau)/r)$ has no $\delta$-singularity.
The insertion of equations \eref{5.10} and \eref{5.11} in equation \eref{5.9a} gives
\beqan 
\fl\Evec^{(3,\,p)}(\rvec,t)&=&\frac{1}{24\pi\eps_0}\left(\nablav^4\vert\vert\frac{\pct^{(3)}(\tau)}{r}
-\frac{1}{c^2}\nablav^2\vert\vert\frac{\ddot{\pct}^{(3)}(\tau)}{r}\right)_{r\ne 0}\nonumber\\
&&+\frac{1}{4\pi\eps_0}\left(\frac{1}{6c^2}\nablav^2\vert\vert\frac{\ddot{\bbox{\Lambda}}(\tau)}{r}-\frac{1}{6c^4}\frac{\qdot{\bbox{\Lambda}}(\tau)}{r}\right)_{r\ne 0}\nonumber\\
&&
-\frac{1}{14\eps_0}\pct^{(3)}(t)\vert\vert\nablav^2\delta(\rvec)
-\frac{1}{2\eps_0}\bbox{\Lambda}(t)\vert\vert\nablav^2\delta(\rvec)+\frac{1}{9\eps_0}\ddot{\Lambda}(t)\delta(\rvec)\ .
\eeqan
From the above equation  we can conclude that $\Evec^{(3,\,p)}(\rvec,t)$ is not invariant to the substitution $\pct^{(3)}\,\to\,\pct^{(3)}$. This substitution introduces an additional term which has precisely the expression of the electric field of  an electric dipole of moment
 \beqa\label{5.13} 
 \bbox{\delta\pvec}^{''}=\frac{1}{6c^2}\ddot{\Lambda}(t)\ .
 \eeqa
Let us express the magnetic field $\Bvec^{(3,\,p)}$:
 \beqa\label{5.14} 
\Bvec^{(3,\,p)}(\rvec,t)=\frac{\mu_0}{24\pi}\nablav\times\left(\nablav^2\vert\vert\frac{\dot{\psft}^{(3)}(\tau)}{r}\right)=\frac{\mu_0}{24\pi}\evec_i\eps_{ijk}\d_j\d_{i_1}\d_{i_2}\frac{\psf_{i_1i_2k}(\tau)}{r}\ .
 \eeqa
The introduction of the {\bf STF} moment $\pct^{(3)}$ has as result 
\beqan
\Bvec^{(3,\,p)}(\rvec,t)=\frac{\mu_0}{24\pi}\nablav\times\left(\nablav^2\vert\vert\frac{\dot{\pct}^{(3)}(\tau)}{r}\right)+\frac{\mu_0}{24\pi}\evec_i\eps_{ijk}\d_j\d_{i_1}\d_{i_2}\frac{\delta_{\{i_1i_2}\lasf_{k\}}(\tau)}{r}
\eeqan
and since $\nablav\times(\nablav^2\vert\vert(\dot{\pct}^{(3)}(\tau)/r))$ has no $\delta$-singularities, we can write
after simple algebraic calculation and after using equation \eref{5.2} that 
 \beqan\label{5.14a} 
\Bvec^{(3,\,p)}(\rvec,t)&=&\left(\frac{\mu_0}{24\pi}\nablav\times\big(\nablav^2\vert\vert\frac{\dot{\pct}^{(3)}(\tau)}{r}+\frac{\mu_0}{4\pi}\;\frac{1}{6c^2}\nablav\times\frac{\tdot{\bbox{\Lambda}}(\tau)}{r}\right)_{r\ne 0}\nonumber\\ &&+\frac{\mu_0}{6c^2}\dot{\bbox{\Lambda}}(t)\times\nablav\delta(\rvec) \ .
 \eeqan
For $r\ne 0$, the additional term introduced by the substitution $\psft^{(3)}\to\pct^{(3)}$ in equation \eref{5.14} represents the magnetic field corresponding to the electric dipole with the moment $\bbox{\delta\pvec}^{''}$ defined by equation \eref{5.13}.\\
Let us consider  the electric moment up to $n=3$, and the magnetic one up to $n=2$ and the corresponding sums of fields   representing one of the first  approximations of a complex system assimilated  with a point-like multipolar system. From the above results, as it was expected, we see that the substitutions of the primitive moments by the corresponding {\bf STF} ones introduce  some additional terms. The sum of these terms is equivalent to the substitution 
\beqan
\pvec\,\to\, \widetilde{\pvec}=\pvec+\delta\pvec\ ,
\eeqan 
where
\beqan
\delta\pvec=\bbox{\delta\pvec}'+\bbox{\delta\pvec}''=-\frac{1}{c^2}\dot{\bbox{t}},
\eeqan
and the vector $\bbox{t}$ is defined as
\beqan
\bbox{t}=\frac{1}{4}\Nsft-\frac{1}{6}\dot{\latens}=\frac{1}{10}\int_{\dom}\rmd^3x\, \left((\rvec.\jvec)\,\rvec-2r^2\,\jvec\right)\ .
\eeqan
This vector is the so-called electric dipolar toroidal moment and represents a first term from a series of toroidal moments introduced by Dubovik {\it et al} \cite{Dubovik-FEC}-\cite{RV} by  generalizing    Zeldovich's original idea that a closed toroidal current represents a certain new kind of dipole   \cite{Zeldovich}. The corresponding $\delta$-singularities can be easily identified in the expressions  established above.\\
The procedure can be continued for the next orders, the technique being the same as in the case of the first orders treated above. Unlike in the static case, in the dynamic one it is more complicated to establish expressions of the singularities for an arbitrary $n$. The recursive character involving a general number of $n$ steps makes such relations hard to derive and apply \cite{cv02}. Furthermore, at the current level of applications, we believe that it is mandatory to have expressions up to $n=3$.

\section{Conclusion}\label{conclusion}
In this paper we have expressed the $\delta$-singularities of the electric and magnetic multipoles. The results were   given for arbitrary multipole orders $n$ in the static case, while in the dynamic one, the outcomes correspond to the lower orders.  However, the calculation algorithm is presented in a manner that   facilitates the processing of the results step by step to the next orders of the dynamic expressions. The central idea of  the article is that, instead of employing the $\delta$-singularities of the functions $f(\tau)/r$, it is more efficient for the higher orders  to search directly such singularities for the multipole fields represented in terms of the {\bf STF} moments. Working in Cartesian coordinates and employing a particular system of parameters and notations adapted to the technique of reducing the primitive Cartesian moments to the corresponding {\bf STF} ones, we have stressed the significance of the first type of moments in the process of searching the field singularities.\\
In our opinion, a first field of application of these results is to extend the classical argumentation of the generalization of the Fermi contact term to arbitrary multipole-multipole interactions, useful in the studies of the hyperfine interaction, \cite{Bucher,Karl}.

\appendix 
\section{Some derivatives of $f(\tau)/r$}\label{A}
\beqan
\d_{i_1}\dots\d_{i_n}\frac{f(\tau)}{r}=\suml^{n}_{l=0}\frac{1}{c^{n-l}r^{l+1}}\;C^{(n,\,l)}_{i_1\dots i_n}
\frac{\rmd^{n-l}f(\tau)}{\rmd t^{n-l}}\ ,\;\;\;\tau=t-\frac{r}{c}\ .
\eeqan
\beqa\label{C0}
C^{(0,0)}=1\ .\;\;
\eeqa
\beqa\label{C1}
C^{(1,0)}_i=-\nu_i\ ,\;\;\;C^{(1,1)}_i=-\nu_i\ .
\eeqa
\beqa\label{C2}
C^{(2,0)}_{ij}=\nu_i\nu_j\ ,\;\;C^{2,1}_{ij}=3\nu_i\nu_j-\delta_{ij}\ ,\;\;\;C^{(2,2)}_{ij}=3\nu_i\nu_j-\delta_{ij}\ .
\eeqa
\beqa\label{C3}
&~&C^{(3,0)}_{ijk}=-\nu_i\nu_j\nu_k\ ,\;\;\;C^{(3,1}_{ijk}=-6\nu_i\nu_j\nu_k+\delta_{\{ij}\nu_{k\}}\ ,\nonumber\\
&~&C^{(3,2)}=-\,15\nu_i\nu_j\nu_k+3\,\delta_{\{ij}\nu_{k\}}\ ,\;\;C^{(3,3)}=-\,15\nu_i\nu_j\nu_k+3\,\delta_{\{ij}\nu_{k\}}\ .
\eeqa
\beqa\label{C4}
 C^{(4,0)}_{ijkl}=\nu_i\nu_j\nu_k\nu_l\ ,\nonumber\\
 C^{(4,1)}_{ijkl}=10\nu_i\nu_j\nu_k\nu_l-\,\delta_{\{ij}\nu_k\nu_{l\}}\ ,\nonumber\\
 C^{(4,2)}_{ijkl}=45\,\nu_i\nu_j\nu_k\nu_l-\,6\delta_{\{ij}\nu_k\nu_{l\}}+\,\delta_{\{ij}\delta_{kl\}}\ ,\nonumber\\
 C^{(4,3)}_{ijkl}=105\,\nu_i\nu_j\nu_k\nu_l-15\,\delta_{\{ij}\nu_k\nu_{l\}}+3\,\delta_{\{ij}\delta_{kl\}}\ ,\nonumber\\
 C^{(4,4)}=105\,\nu_i\nu_j\nu_k\nu_l-15\,\delta_{\{ij}\nu_k\nu_{l\}}+3\,\delta_{\{ij}\delta_{kl\}}\ .
\eeqa

\section{Some general formulas for {\bf STF} tensors singularities}\label{B}
When searching the $\delta$-singularities of the electromagnetic  fields,  we have to calculate  singularities of some derivatives of {\bf STF} electric  and magnetic moments as
\beqa\label{D1}
\nablav^{n+1}\vert\vert\frac{\Tsft^{(n)}(\tau)}{r},\;\;\nablav\times\left(\nablav^{n-1}\vert\vert\frac{\Tsft^{(n)}(\tau)}{r}\right),
\eeqa
where $\Tsft^{(n)}$ is one of the {\bf STF} electric or magnetic moments.\\
Let us search the $\delta$-singularities of the first expression for $n\,\ge\,2$ starting from the definition
\beqa\label{D2}
\fl\left\langle\,\left(\nablav^{n+1}\vert\vert\frac{\Tsft^{(n)}(\tau)}{r}\right)_{(0)},\;\phi\right\rangle&=&
\lim_{\eps\to0}\evec_i\left[\oint_{\Sigma_\eps}\rmd S\,\nu_i\d_{i_1}\dots\d_{i_n}\frac{\Tsf_{i_1\dots i_n}(\tau)}{r}\,\phi(\rvec)\right.\nonumber\\
\fl&&-\left.\int_{\dom_\eps}\rmd^3x\,\d_{i_1}\dots\d_{i_n}\frac{\Tsf_{i_1\dots i_n}(\tau)}{r}\,\phi(\rvec)\,\d_i\phi(\rvec)\right]\ .
\eeqa
Denoting  $\bbox{L}_\sigma$ the limit corresponding to the surface integral from the last equation,  the  insertion of equations \eref{a.19}, \eref{a.24} gives:
\beqa\label{D4}
\fl\;\;\;\;\;\;\;\;\;\;\;\;\bbox{L}_\sigma&=&\lim_{\eps\to0}\evec_i\,\suml^n_{l=0}\suml_{\al\ge0}\suml_{\la\ge0}\frac{(-1)^\la\eps^{\al-l+1+\la}}{c^{n-l+\la}\al!\la!}\nonumber\\
\fl\;\;\;\;\;\;\;\;\;\;\;\;&\times&\oint_{\Sigma_\eps}\rmd\Omega(\nuvec)\,\nu_i\,C^{(n,\,l)}_{i_1\dots i_n}\nu_i\nu_{j_1}\dots\nu_{j_\al}
\frac{\rmd^{n-l+\la}}{\rmd t^{n-l+\la}}\Tsf_{i_1\dots i_n}(t)\left(\d_{j_1}\dots\d_{j_\al}\,\phi\right)_0\ .
\eeqa
Considering the expressions \eref{a.20} of the coefficients $C^{(n,\,l)}$, in equation \eref{D4} we can perform the substitutions 
\beqan
C^{(n,\,l)}_{i_1\dots i_n}\,\to\,K^{(n,\,l)}_0\,\nu_{i_1}\dots \nu_{i_n}
\eeqan
without changing the result since the contractions of $\Tsf_{i_1\dots i_n}$ with all the terms from the coefficients $C^{(n,\,l)}_{i_1\dots i_n}$ containing at least a symbol Kronecker vanish. Expressing the integral in this equation as angular average, we can write
\beqa\label{D5}
\bbox{L}_\sigma&=&4\pi\,\evec_i\lim_{\eps\to0}\suml^n_{l=0}\suml_{\al\ge0}\suml_{\la\ge0}\frac{(-1)^\la\eps^{\al-l+1+\la}K^{(n,\,l}_0}{c^{n-l+\la}\al!\la!}\nonumber\\
&\times&\left\langle\,\nu_{i_1}\dots\nu_{i_n}\nu_i\,\nu_{j_1}\dots \nu_{j_\al}\right\rangle\frac{\rmd^{n-l+\la}}{\rmd t^{n-l+\la}}\Tsf_{i_1\dots i_n}(t)\left(\d_{j_1}\dots\d_{j_\al}\,\phi\right)_0\ .
\eeqa
As we can see from equation \eref{3.6}, the contractions 
\beqan
\left\langle\,\nu_{i_1}\dots\nu_{i_n}\nu_i\,\nu_{j_1}\dots \nu_{j_\al}\right\rangle\,\Tsf_{i_1\dots i_n}(t)
\eeqan
are different from zero only if $\al+1+n\,\ge\,2n$, i.e.\
\beqa\label{D6}
\al\,\ge\,n-1\ .
\eeqa
Let $e=\al-l+1+\la$ be the power of $\eps$ in equation \eref{D5}. Since $l\,\le\,n$ and $\la\,\ge\,0$, the inequality \eref {D6} implies $e\,\ge\,0$, the equality holding only for $\al=n-1,\;l=n,\,\la=0$. $\bbox{L}_\sigma\ne 0$ only in this case since for $e\,>\,0$ the limit vanishes because of the positive values of the powers of $\eps$. Therefore, equation \eref{D5} becomes
\beqa\label{D7}
\fl\;\;\;\;\;\;\;\;\;\;\;\;\bbox{L}_\sigma=4\pi\,\evec_i\,\frac{K^{(n,\,n)}_0}{(n-1)!}\left\langle\,\nu_{i_1}\dots\nu_{i_n}\,\nu_i\,\nu_{j_1}\dots\nu_{j_{n-1}}\right\rangle\Tsf_{i_1\dots i_n}(t)\,\left(\d_{j_1}\dots\d_{j_{n-1}}\,\phi\right)_0\ .
\eeqa
As one can easily see  from equation \eref{3.6}, the remaining contractions are different from zero only for the $n!$ terms of the form $\delta_{i_1j_1}\dots\delta_{i_{n-1}j_{n-1}}\delta_{i_ni}$, and, inserting the value \eref{a.22} of $K^{(n,\,n)}_0$ we can finally write 
\beqa\label{D8}
\fl\;\bbox{L}_\sigma=\evec_i \frac{(-1)^n4\pi\,n}{2n+1}\Tsf_{i_1\dots i_{n-1}\,i}(t)\left(\d_{i_1}\dots\d_{i_{n-1}}\,\phi\right)_0
=\frac{(-1)^n4\pi\,n}{2n+1}\;\Tsft^{(n)}(t)\vert\vert\left(\nablav^{n-1}\phi\right)_0.
\eeqa
The volume integral limit from equation \eref{D2} can be processed applying repeatedly the Gauss theorem. The new surface integral limit can be treated as in the previous case since it is represented by an equation similar to equation \eref{D5} with $n\,\to\,n-1$.  The contractions to be considered  this time are
\beqan
\left\langle\,\nu_{i_1}\dots\nu_{i_n}\,\nu_{j_1}\dots\nu_{j_\al}\right\rangle\,\Tsf_{i_1\dots i_n}\ .
\eeqan
They are different from zero  only if $\al\,\ge\,n$ and, since $l\,\le\,n-1$, the exponents  of $\eps$ are $e=\al-l+1+\al\,\ge\,2$ such that the limit of this integral for $\eps\to0$  cancels.  Therefore, equation \eref{D8} inserted in equation \eref{D2} has as final result 
\beqa\label{D9}
\left(\nablav^{n+1}\vert\vert\frac{\Tsft^{(n)}(\tau)}{r}\right)_{(0)}=4\pi\,\frac{(-1)^n\,n}{2n+1}\,\,\Tsft^{(n)}(t)\vert\vert\nablav^{n-1}\delta(\rvec)\ .
\eeqa
The curl expression from equation \eref{D1} has no $\delta$-singularities and this result is obtained applying  the same  procedure as in the case of the first expression from this equation. Indeed, processing the corresponding surface integral limit,  we can reduce it to  terms having as factors the following contractions  
\beqan
\eps_{ijk}\left\langle\,\nu_{i_1}\dots\nu_{i_{n-1}}\,\nu_j\,\nu_{j_1}\dots\nu_{j_{n-2}}\right\rangle\,\Tsf_{i_1\dots i_{n-1}\,k}(t)\ .
\eeqan
But
\beqan
\eps_{ijk}\left\langle\,\nu_{i_1}\dots\nu_{i_{n-1}}\,\nu_j\,\nu_{j_1}\dots\nu_{j_{n-2}}\right\rangle\,\Tsf_{i_1\dots i_{n-1}\,k}(t)\sim \eps_{ijk}\Tsf_{i_1\dots i_{n-2}\,jk}=0\ ,
\eeqan
therefore,
\beqa\label{D10}
\left[\nablav\times\left(\nablav^{n-1}\vert\vert\frac{\Tsft^{(n)}(\tau)}{r}\right)\right]_{(0)}=0\ .
\eeqa
\vspace{1.cm}\\
{\it Acknowledgments}\\
The authors would like to thank Dr.\ Raluca Paiu for her help with the documentation. The work of RZ was supported by grant ID946 (no.\ 44/2007) of Romanian National Authority for Scientific Research.

\end{document}